# Lift Enhancement by Dynamically Changing Wingspan in Forward Flapping Flight

Shizhao Wang[1], Xing Zhang[1], Guowei He[1a)], Tianshu Liu[2,1]

(09/10/2013)

[1]The State Key Laboratory of Nonlinear Mechanics, Institute of Mechanics, Chinese Academy of Sciences, Beijing 100190, P. R. China

[2]Department of Mechanical and Aerospace Engineering, Western Michigan University, Kalamazoo, MI 49008, USA

---

[a)] The corresponding author, Email: hgw@lnm.imech.ac.cn, Telephone: 86-10-82543969

**ABSTRACT**

Stretching and retracting wingspan has been widely observed in the flight of birds and bats, and its effects on the aerodynamic performance particularly lift generation are intriguing. The rectangular flat-plate flapping wing with a sinusoidally stretching and retracting wingspan is proposed as a simple model of biologically-inspired dynamic morphing wings. Direct numerical simulations of the low-Reynolds-number flows around the flapping morphing wing in a parametric space are conducted by using immersed boundary method. It is found that the instantaneous and time-averaged lift coefficients of the wing can be significantly enhanced by dynamically changing wingspan in a flapping cycle. The lift enhancement is caused not only by changing the lifting surface area, but also manipulating the flow structures that are responsible to the generation of the vortex lift. The physical mechanisms behind the lift enhancement are explored by examining the three-dimensional flow structures around the flapping wing.

# I. INTRODUCTION

Animal flight has provided inspirations for new designs of Micro Aerial Vehicles (MAVs) since natural flyers usually have the extraordinary maneuvering capability, high lift and efficient propulsion. On the other hand, the complex low Reynolds number unsteady aerodynamics in flapping flight poses challenges to researchers in aerodynamics and fluid mechanics because the large body of the knowledge and database of the design of large fixed-wing aircraft is not suitable in the design of such MAVs. Therefore, it is highly desirable to understand the physical mechanisms of lift generation in animal flight. Insects, birds and bats are three groups of extant flying animals. Notwithstanding the different morphology and kinematics of their wings, the high-lift-generating mechanisms of flapping wings at low Reynolds numbers is one of the common merits of these flyers. Some unique mechanisms to enhance lift particularly in insect flight have been identified, including the 'clap and fling' mechanism [1, 2], stable attached leading-edge vortex (LEV) during dynamic stall [3-7], rapid acceleration of the wing at the beginning of a stroke, fast pitching-up rotation of the wing near the end of the stroke [8], wake capture mechanism [9, 10] and wing-body/wing interaction [11-13].

The aerodynamic forces acting on flapping wings are directly related to flow structures around the wings. The flow structures near wings and in the wakes have been investigated extensively by measuring flows around a natural flyer or a mechanical model [3, 14-19] and by numerical simulations [4]. Flow structures around a heaving and/or pitching rigid wing with several planforms as a simplified flapping wing have been also investigated [20-25]. These studies on the high-lift mechanisms and flow structures of flapping wings at low Reynolds numbers provide understandings into the aerodynamics of low Reynolds number unsteady flows of animal flight. The comprehensive reviews on this area are given in Ref. 26-30. Recently, the concepts of flexible wings have been inspired by animal flight. Flexible wings of natural flyers could be adopted to improve the aerodynamic performance of MAVs. Insects usually have flexible membrane wings that deform elastically during the flapping, modifying the flow around the wing. Birds have flexible feathers, and bats have elastic membrane wings. The effects of the wing flexibility have attracted attentions of researchers [31-36].

More significantly, in contrast to insects, the wingspan and wing planform are actively changed though skeleton motions in flapping flight of birds and bats [16-18, 37]. This dynamic wing morphing is ubiquitous in bird and bat flight. For example, a bat wing has a bone skeleton with more than ten joints and therefore a bat can quickly change the wingspan and planform by moving joints in a controllable way. In general, the wing stretches outward in the downstroke and retracts inward to the body in the upstroke. The ratio between the minimum and the maximum wingspans of a Pallas' long tongued bat could be as low as 0.6 [38]. The current research of morphing wings has been limited on steady and quasi-steady morphing to meet mission requirements at different flight regimes (such as takeoff, landing and cruising) by using smart materials [39]. Therefore, it is highly desirable to investigate the dynamic morphing of a flapping wing for potential improvement of flight performance.

For a dynamic morphing flapping wing with the time-dependent wing area, the lift coefficient for forward flight is defined as

$$Cl(t) = F_z(t) / q_\infty S(t), \tag{1}$$

where the $F_z$ is the lift acting on the wing, $q_\infty = 0.5\rho U_\infty^2$ is the dynamical pressure, $U_\infty$ is the freestream velocity (or forward flight velocity), and $S(t)$ is the instantaneous wing area. Theoretically, the effect of changing the wing area is removed in $Cl(t)$  The time-averaged lift over a flapping period $T$ is given by

$$\langle F_z \rangle_T = q_\infty T^{-1} \int_0^T Cl(t) S(t) dt, \tag{2}$$

where $\langle \bullet \rangle_T = T^{-1} \int_0^T \bullet \, dt$ is the time-averaging operator. According to Eq. (2), even when the time-averaged lift coefficient is zero, i.e., $\langle Cl \rangle_T = 0$, the positive time-averaged lift ($\langle F_z \rangle_T > 0$) could be still generated by dynamically changing the wing area in flapping flight such that there is a positive correlation $\langle Cl(t)S(t) \rangle_T > 0$. Bird and bat flight is just a good example in which the wing area is increased in the downstroke and decreased in the upstroke.

It is clear that the effect of a changing lifting surface area will directly alter the lift generation. Nevertheless, a question is whether an increase of the lift coefficient $Cl(t)$ could be achieved by manipulating flow structures induced by dynamically changing the wing area in in addition to the effect of changing area. There are few studies on complex flows around a flapping wing with large dynamic morphing like a bat or bird wing. The changes of the flow structures induced by the dynamic morphing and their effects on the lift coefficient are not well understood. To gain a better understanding into this problem, a canonical morphing flapping wing, a rectangular flat-plate wing with a stretching and retracting wingspan, is considered. This simplified model characterizes the main spanwise morphing features of flapping bird and bat wings although wings of various birds and bats have more complex planforms and kinematics. The fundamental question is how such stretching and retracting wingspan during a wing flapping cycle affects flow structures and changes the lift coefficient $Cl(t)$ as a result.

The objective of this work is to demonstrate through direct numerical simulations that the dynamic morphing of a flapping wing can enhance the lift and explore the physical mechanisms behind the lift enhancement. The strategy of solving this problem is briefly outlined and the structure of the paper is described as follows. First, a generic dynamic morphing flapping wing model is proposed, which captures the main morphing feature of a typical bird or bat wing in terms of stretching and retracting wingspan during flapping. This model is a rectangular flat-plate with a sinusoidally varying wingspan that reaches the maximum during the downstroke and minimum during the upstroke. The wing geometry and kinematics in a parametric space are given and the relevant aerodynamic parameters are defined to describe the lift enhancement in Section II. A concise description of the numerical method and setting is given in Section II. In Section III, the enhancement of the lift coefficient is examined in the parametric space and it is indicated that the lift can be significantly enhanced by altering the flow structures associated with stretching and retracting the wingspan besides the effect of changing the wing area. Further, the vortical structures that are responsible to the enhancement of the lift coefficient due to stretching and retracting wingspan are identified in Section IV. It is observed that the leading-edge vortices on the upper surface in the upstroke are significantly intensified and the shorter and weaker leading-edge vortices occur on the lower surface in both the upstroke and downstroke. These structures lead to the overall vortex lift enhancement. These observations are further supported by a data analysis based on the lift decomposition into the Lamb vector term (the vortex lift) and the local acceleration term in Section V. Finally, the conclusions are drawn in Section VI. The effect of the dynamic wing aspect ratio is discussed based on a quasi-steady lifting line model in Appendix A and the induced drag associated with the enhanced lift is discussed in Appendix B.

## II. MODEL AND METHOD

### A. Generic morphing flapping wing

The geometry and kinematics of wings of flying birds and bats are complicated [16-18, 37]. Therefore, to gain a clear understanding of the aerodynamics of a morphing flapping wing, the wing geometry and kinematics should be suitably simplified while the main morphing feature of the wing is retained. A generic morphing flapping wing is proposed, as illustrated in Fig. 1, which characterizes the dynamic change of the wing area by stretching and retracting the wingspan. A rectangular flat-plate wing with a constant geometrical angle of attack (AoA) $\alpha$ heaves harmonically in a uniform freestream flow. Meanwhile, its wingspan stretches outward before reaching a certain position in the downstroke and retracts inward to the center line before reaching a certain position in the upstroke. The wing has zero thickness. The wing chord is uniform and remains constant during flapping. The flapping kinematics is described in a fixed laboratory coordinate system, as shown in Fig. 1. The x-axis points downstream in the direction of the freestream flow, the y-axis is in the spanwise direction, and the z-axis is in vertical direction pointing upward. The flapping kinematics of the center of the wing is prescribed by

$$z_w = A\sin(2\pi f t), \quad (3)$$

where $z_w$ is the vertical position of the wing center, $A$ is the heaving amplitude, $f$ is the flapping frequency. The flapping Strouhal number is defined as $St = 2fA/U_\infty$. The time history of $z_w$ is shown in Fig. 2, where $T^* = t/T - 1/4 = ft - 1/4$ is a non-dimensional time in which the time is shifted by $1/4$ of the period so that the downstroke in the flapping motion given by Eq. (3) starts at $T^* = 0, 1, 2, \cdots$ when $z_w$ reaches the maximum.

The spanwise stretching and retracting motion of the wing has the same frequency as that of flapping. The wingspan $L$ reaches the maximum during the downstroke and minimum during the upstroke. The wing aspect ratio is prescribed as a function of time, i.e.,

$$AR = L/c = AR_0(a - b\sin(2\pi f t + \phi)), \quad (4)$$

where $AR_0 = L_o/c$ is the characteristic aspect ratio, $L_o$ is the characteristic wingspan, $c$ is the constant chord, $a$ and $b$ are the coefficients that specify the stretching and retracting amplitude, and $\phi$ is the phase difference between the flapping and stretching/retracting motions. The span ratio, which is defined as the ratio between the minimum and maximum wingspans ($L_{min}$ and $L_{max}$), is introduced to measure the magnitude of stretching and retracting wingspan, i.e.,

$$SR = L_{min} / L_{max}. \tag{5}$$

The span ratio depends on $a$ and $b$ for a given value of $AR_0$, as shown in Table 1. The wing morphing is characterized by the span ratio. The time history of the aspect ratio for $SR = 0.5$ and $\phi = \pi/2$ is shown in Fig. 2 along with the history of the flapping motion. When the non-dimensional variables as $z_w^* = z_w/c$, $A^* = A/c$, $f^* = fc/U_\infty$, $t^* = tU_\infty/c$, $L_0^* = L_0/c$ are used, the non-dimensional forms of Eqs. (3) and (4) remain the same, which will be used in the following sections. In summary, there are the four kinematical parameters: the span ratio $SR$, the phase difference $\phi$, the flapping Strouhal number $St = 2fA/U_\infty$, and the relative flapping magnitude $A^* = A/c$. The geometrical angle of attack (AoA) $\alpha$ and the characteristic aspect ratio $AR_0$ are the other relevant parameters are the other relevant parameter. The wing area changes as a function of time when the wingspan is stretched and retracted during a flapping cycle. The coefficients $a$ and $b$ in Eq. (4) are selected such that $L_0/c = L_{max}/c = 4$ for all the cases studied in the present work. The wing area changes as a function of time when the wingspan is stretched and retracted during a flapping cycle, and the instantaneous wing area is $S(t) = L(t)c$.

## B. Aerodynamic parameters

The lift coefficient $Cl(t)$ is a function of the parameters $SR$, $\phi$, $St$, $A^*$ and $\alpha$. To compare the wings with the dynamically changing wingspan and the fixed wingspan, the instantaneous increment of $Cl(t)$ is introduced, i.e.,

$$\Delta Cl(t) = Cl(t) - Cl(t; SR = 1), \tag{6}$$

where $Cl(t)$ and $Cl(t; SR=1)$ are the lift coefficients based on the dynamically changing wingspan and the fixed wingspan, respectively, while the other parameters remain the same. Essentially, $\Delta Cl(t)$ represents the lift increment generated by the fluid-mechanic effect induced by the dynamic morphing. Furthermore, the time-averaged quantity $\langle \Delta Cl \rangle_T$ is used as a measure of the enhancement of the lift coefficient in the parametric space $(SR, \phi, St, A^*, \alpha)$.

For comparison with $Cl(t)$, it is also useful to characterize the overall lift generation by both the effect of changing the wingspan and the fluid-mechanic effect induced by dynamic morphing. For this purpose, the lift coefficient based on the maximum wing area is introduced, i.e.,

$$Cl_{S\_max}(t) = F_z(t) / q_\infty S_{max}, \tag{7}$$

where $S_{max}$ is the maximum wing area (in this case $S_{max} = 4c^2$). Similar to Eq. (6), the increment of is

$$\Delta Cl_{S\_max}(t) = Cl_{S\_max}(t) - Cl_{S\_max}(t; SR = 1), \tag{8}$$

and the time-averaged one is $\langle \Delta Cl_{S\_max} \rangle_T$. For $SR = 1$, $Cl(t; SR=1) = Cl_{S\_max}(t; SR=1)$ since the area for normalization is $S_{max}$. Introducing a correlation coefficient $C_{FS} = \langle F_z(t) S^{-1}(t) \rangle_T \langle F_z \rangle_T^{-1} \langle S \rangle_T$, based on the definitions Eqs. (1) and (7), we have a relation $\langle Cl \rangle_T = C_{FS} \langle Cl_{S\_max} \rangle_T S_{max} \langle S \rangle_T^{-1}$.

According to the kinematical equation Eq. (4) for the wingspan, we know $\langle S \rangle_T = S_{max} a/(a+b)$. Therefore, there are the proportional relations $\langle Cl \rangle_T = C_{FS} \langle Cl_{S\_max} \rangle_T (a+b)/a$ and $\langle \Delta Cl \rangle_T = C_{FS} \langle \Delta Cl_{S\_max} \rangle_T (a+b)/a$. In general, the correlation coefficient $C_{FS}$ is a function of the parameters $(SR, \phi, St, A^*, \alpha)$.

### C. Numerical method and settings

The flow around the flapping rectangular wing is governed by the incompressible Navier-Stokes (NS) equations

$$\begin{aligned} \nabla \cdot \boldsymbol{u} &= 0 \\ \frac{\partial \boldsymbol{u}}{\partial t} + \boldsymbol{u} \cdot \nabla \boldsymbol{u} &= -\nabla p + \frac{1}{Re} \nabla^2 \boldsymbol{u} + \boldsymbol{f} \end{aligned} \quad (9)$$

where $\boldsymbol{u}$ is the non-dimensional velocity normalized by $U_\infty$, $p$ is the non-dimensional pressure normalized by $\rho U_\infty^2$, $\boldsymbol{f}$ is the non-dimensional body force, and $Re = U_\infty c / \nu$ is the Reynolds number. The unsteady flow with a moving boundary is handled by using a semi-implicit Immersed Boundary (IB) method in the frame work of discrete stream function formula [40]. With this method, the geometry and kinematics of the flapping wing are described by a set of the Lagrangian grid points. The NS equations are solved on a Cartesian grid by using the discrete stream function (or exact projection) approach, in which the divergence-free condition is exactly satisfied.

This approach is briefly described as follows. The discrete form of Eq. (9) is

$$\begin{bmatrix} \mathbf{A} & \mathbf{G} \\ \mathbf{D} & 0 \end{bmatrix} \begin{bmatrix} q^{n+1} \\ p \end{bmatrix} = \begin{bmatrix} r^n \\ 0 \end{bmatrix} - \begin{bmatrix} bc_1 \\ bc_2 \end{bmatrix} + \begin{bmatrix} f \\ 0 \end{bmatrix}, \quad (10)$$

where $q^{n+1}$ is the discrete velocity flux at time $n+1$, $p$ and $f$ are the discrete pressure and body force, respectively, $r^n$ is the explicit right-hand side of the momentum equation, $bc_1$ and $bc_2$ are the boundary conditions, and $\mathbf{A}$, $\mathbf{G}$ and $\mathbf{D}$ are the implicit operator, gradient operator and divergence operator, respectively. The matrix $\mathbf{C}$ and $\mathbf{R}$ are constructed as a null space of $\mathbf{D}$ and $\mathbf{G}$, respectively, i.e. $\mathbf{DC}=0$ and $\mathbf{GR}=0$. With this definition, the discrete equation Eq. (10) is reduced to

$$\mathbf{RAC}s = \mathbf{R}r^n - \mathbf{R}bc_1 + \mathbf{R}f, \qquad (11)$$

where $s$ is defined by $q^{n+1}=\mathbf{D}s$, which can be regarded as a discrete stream-function [41]. The effect of the flapping wing on the flow is represented by adding a body force term to the momentum equations [42, 43]. In order to impose the non-slip boundary condition on the flapping wing, the forcing term is calculated implicitly by solving a linear system regarding the interpolation between the Lagrangian and Eulerian grid points, i.e.,

$$\sum_{j=1}^{M}\left(\sum_{\mathbf{x}}\delta_h(\mathbf{x}-\mathbf{X}_j)\delta_h(\mathbf{x}-\mathbf{X}_k)(\delta s)^2(\delta h)^3\right)\mathbf{F}(\mathbf{X}_j) = \frac{\mathbf{U}^{n+1}(\mathbf{X}_k)-\mathbf{U}^*(\mathbf{X}_k)}{\Delta t}, \qquad (12)$$

where $\mathbf{x}$ and $\mathbf{X}$ are the position vectors of the Eulerian and Lagrangian grid points, respectively, $\mathbf{F}$ is the body force at the Lagrangian grid points, $\mathbf{U}^{n+1}(\mathbf{X}_k)$ and $\mathbf{U}^*(\mathbf{X}_k)$ are the desired velocity and predicted velocity at the $kth$ Lagrangian grid point, respectively, $\delta_h$ is the discrete Delta function [the form provided by Peskin [42] is used in the present simulations], $\delta h$ and $\delta s$ are the Eulerian and Lagrangian grid size, respectively, and $M$ is the number of the Lagrangian grid points. The body forces on the Eulerian grid points are calculated by

$$\mathbf{f}(\mathbf{x}) = \sum_{j=1}^{M} \mathbf{F}(\mathbf{X}_j) \delta_h (\mathbf{x} - \mathbf{X}_j)(\delta s)^2. \tag{13}$$

The detailed description of the numerical method is given by Wang and Zhang [40].

The present work focus on the low Reynolds number flows around a flapping wing. The Reynolds number $Re = U_\infty c / \nu$ is fixed at 300 for all the cases. The characteristic aspect ratio is fixed at $AR_0 = 4$. The flapping rectangular wing with $SR = 0.5$, $St = 0.3$, $A^* = 0.25$, and $\alpha = 0°$ is selected as a typical case. The parametric effects on the lift enhancement are investigated. A non-dimensional computational domain of $[-16,34] \times [-15,15] \times [-15,15]$ in the streamwise, spanwise, and vertical directions is used. The uniform freestream flow is specified at the inlet of the computational domain. The free convection boundary condition is set at the outlet. The non-slip boundary condition is satisfied on the surface of the flapping wing. The symmetric boundary conditions are imposed on the other boundaries. The flow is uniform with $(1,0,0)$ at $t = 0$, and the flapping wing appears at $t = 0^+$.

An unstructured Cartesian grid, with local refinement using hanging nodes, is used to discretize the computational domain. The total number of discrete cells is 10,189,700 with the grid size $\delta h / c = 0.02$ in a refined domain of $[-1,1] \times [-3,3] \times [-1,1]$ around the flapping wing, $\delta h / c = 0.04$ in a region of $[-2,8] \times [-4,4] \times [-2,2]$ to resolve the flow structures in the wake, and $\delta h / c \leq 0.32$ on the far field boundaries. The Lagrangian grid size varies with time in a range from the minimum of $\delta s / c = 0.01$ to the maximum of $\delta s / c = 0.02$. Figure 3 shows the time histories of the lift coefficient obtained in DNS with different minimum grid sizes and different time steps in the typical case of $SR=0.5$, $\phi=\pi/2$, $St=0.3$, $A^*=0.25$ and $\alpha=0°$. The reasonable converged results are achieved in this work when the minimum grid size of $\delta h/c=0.02$ is used as shown in Fig. 3(a). The independency of the results on the time step is shown in Fig. 3(b). The time step of $\Delta t^* < 0.005$ is selected to keep the maximum Courant–Friedrichs–Lewy (CFL) number being 0.5. The validation of the present code in various unsteady flows is described by Wang and Zhang [40].

TABLE 1. Coefficients specifying the amplitude of stretching and retracting wingspan

| SR    | a      | b      | $AR_0$ |
|-------|--------|--------|--------|
| 0.25  | 0.625  | 0.375  | 4.0    |
| 0.375 | 0.6875 | 0.3125 | 4.0    |
| 0.5   | 0.75   | 0.25   | 4.0    |
| 0.6   | 0.8    | 0.2    | 4.0    |
| 0.7   | 0.85   | 0.15   | 4.0    |
| 0.75  | 0.875  | 0.125  | 4.0    |
| 0.8   | 0.9    | 0.1    | 4.0    |
| 0.85  | 0.925  | 0.075  | 4.0    |
| 0.9   | 0.95   | 0.05   | 4.0    |
| 1.0   | 1.0    | 0.0    | 4.0    |

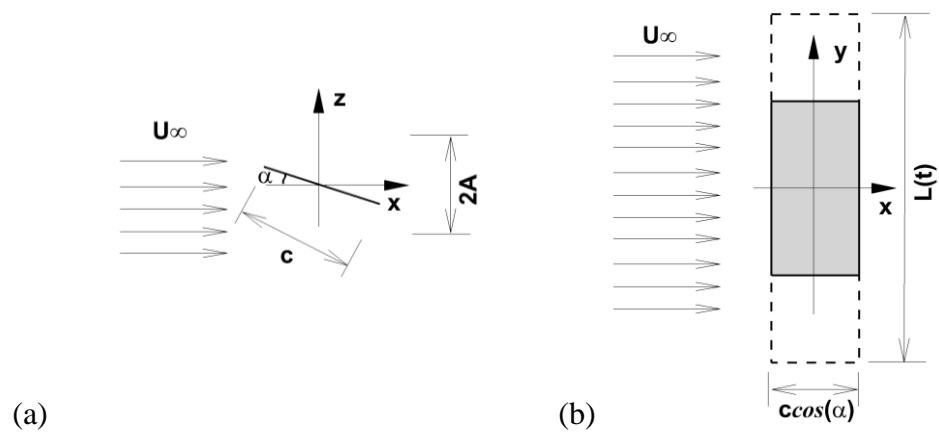

(a)          (b)

FIG. 1. Schematic of the computational model: (a) side view and (b) top view.

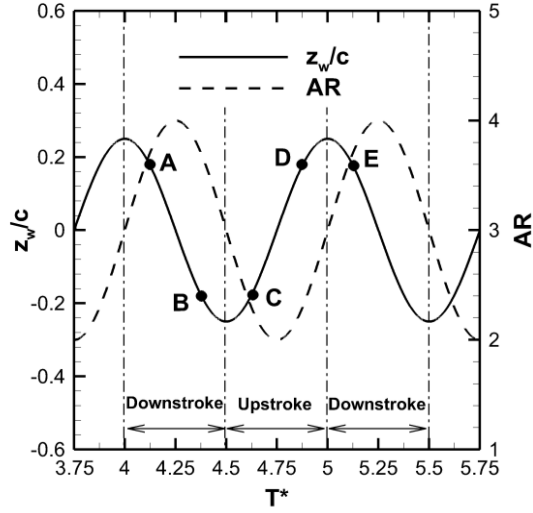

FIG. 2. The time histories of the vertical displacement of the wing and the aspect ratio in the case of $SR = 0.5$. The solid circles A, B, C, D and E denote five key moments $T^* = 4.125$, $4.375$, $4.625$, $4.875$ and $5.125$, respectively, for illustration of the flow structures in Section IV.

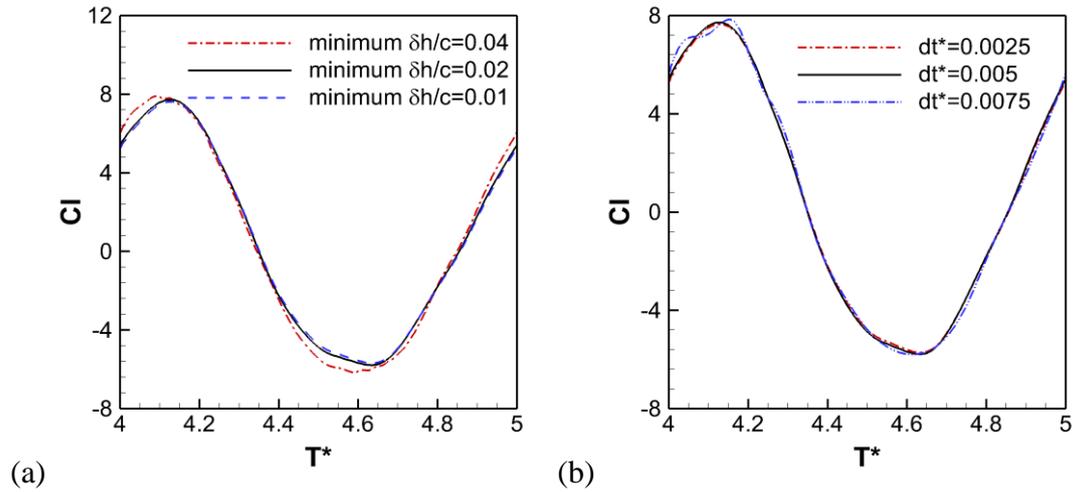

(a)  (b)

FIG. 3. The time histories of the lift coefficient obtained in DNS with (a) different minimum grid sizes and (b) different time steps in the typical case of $SR = 0.5$, $\phi = \pi/2$, $St = 0.3$, $A^* = 0.25$ and $\alpha = 0°$.

## III. LIFT ENHANCEMENT

The lift acting on a flapping wing with a stretching and retracting wingspan depends on the span ratio ($SR$), phase difference ($\phi$), Strouhal number ($St$), relative flapping amplitude ($A^*$) and geometrical AoA ($\alpha$). To examine the lift enhancement, the lift coefficients and their increments are evaluated in the parametric space $(SR, \phi, St, A^*, \alpha)$. The effect of the span ratio ($SR$) on the lift enhancement is first investigated in Section III. A and then the effects of other parameters $(\phi, St, A^*, \alpha)$ are examined in Section III B.

### A. Effect of span ratio

We consider a typical case of a flapping wing with a stretching and retracting wingspan, where the kinematical parameters are $(SR, \phi, St, A^*, \alpha) = (0.5, \pi/2, 0.3, 0.25, 0^o)$ (referred to as the case of $SR = 0.5$). For comparison, the corresponding wing with the fixed wingspan in a flapping cycle are taken as $(SR, \phi, St, A^*, \alpha) = (1.0, \pi/2, 0.3, 0.25, 0^o)$ (referred to as the case of $SR = 1.0$). In this case of $SR = 1.0$, the lift coefficient $Cl(t)$ defined by Eq. (1) is the same as $Cl_{S\_max}(t)$ defined by Eq. (7). Figure 4(a) shows the time histories of $Cl(t)$ and $Cl_{S\_max}(t)$ in the cases of $SR = 0.5$ and $SR = 1.0$. Overall, both the lift coefficients $Cl(t)$ and $Cl_{S\_max}(t)$ for $SR = 0.5$ are larger than those for $SR = 1.0$ during the most time of a flapping cycle. This observation indicates that the stretching and retracting wingspan motion can increase the total lift or the time-averaged lift. Further, since the instantaneous lift coefficient $Cl(t)$ removes the effect of the changing wing area, the fact that $Cl(t)$ for $SR = 0.5$ is larger than that for $SR = 1.0$ leads to an important consequence that the fluid-mechanic mechanism may be responsible to the lift enhancement in addition to the effect of changing the lifting surface area.

In the case of $SR = 0.5$, the maximum of $Cl_{S\_max}(t)$ is reached when the wing is at about one-third of the downstroke. At the beginning of the upstroke, the wing has the minimum of $Cl_{S\_max}(t)$. The maximum magnitude of $Cl_{S\_max}(t)$ generated in the downstroke is about two times of the maximum magnitude of the negative one in the upstroke. This asymmetric lift generation in the downstroke and upstroke results in $\langle Cl_{S\_max} \rangle_T > 0$; In the case of $SR = 1.0$ where the wingspan remains constant during the flapping, $Cl_{S\_max}(t)$ reaches the maximum when the wing moves about one-third of the downstoke. The lift generation during the downstroke is similar to that in the $SR=0.5$ case, except that the maximum $Cl_{S\_max}(t)$ is about 4% smaller. However, unlike the $SR=0.5$ case, the minimum $Cl_{S\_max}(t)$ is reached when the wing is at about one-third of the upstroke. The peak magnitude of the negative $Cl_{S\_max}(t)$ in the upstroke is equals to that of the positive $Cl_{S\_max}(t)$ in the downstroke. In contrast to the case of $SR = 0.5$, this symmetric heaving kinematics without stretching and retracting wingspan in the downstroke and upstroke results in $\langle Cl_{S\_max} \rangle_T = 0$.

Figure 4(b) shows the time history of the lift coefficient increment $\Delta Cl(t)$ defined by Eq. (6). It is found that $\Delta Cl(t)$) is positive in the most time of the upstroke and downstroke and there are two peaks in this period. The fact that $\Delta Cl(t)$ is positive indicates that the additional lift enhancement is achieved by the fluid-mechanic mechanism induced by stretching and retracting wingspan. Figure 4(c) shows the time history of the increment $\Delta Cl_{S\_max}(t)$. It can be seen that the flapping wing generates the large positive lift increment ($\Delta Cl_{S\_max} > 0$) in about 2/3 of the upstroke period for $SR = 0.5$. As expected, this is mainly caused by both retracting the wingspan during the upstroke and modifying the vortical structures associated with the spanwise motion. During the most time of the downstroke, $\Delta Cl_{S\_max}(t)$ remains positive although its value is relatively small.

Furthermore, we evaluate the time-averaged lift coefficients and find that $\langle Cl_{S\_max}\rangle_T = \langle Cl\rangle_T = 0$ for $SR = 1.0$, $\langle Cl_{S\_max}\rangle_T = 0.81$ and $\langle Cl\rangle_T = 0.42$ for $SR = 0.5$. The time-averaged lift coefficient $\langle Cl\rangle_T = 0.42$ for $SR = 0.5$ is significantly larger than $\langle Cl\rangle_T = 0$ for $SR = 1.0$ and is about half of $\langle Cl_{S\_max}\rangle_T = 0.81$ for $SR = 0.5$. The time-averaged lift coefficient $\langle Cl\rangle_T$ mainly represents the contribution to the lift enhancement by the altered flow structures associated with stretching and retracting wingspan. This comparison further confirms that the fluid-mechanic mechanism associated with stretching and retracting wingspan indeed makes a significant contribution to the lift enhancement that is comparable to that generated by changing the lift surface area. Figure 5 shows the time-averaged increment of the lift coefficient $\langle \Delta Cl\rangle_T$ as a function of $SR$ for $\phi = \pi/2$, $St = 0.3$, $A^* = 0.25$, and $\alpha = 0^o$. The value of $\langle \Delta Cl\rangle_T$ remains positive although it monotonically decays as $SR$ increases, indicating that the lift enhancement can be achieved by the fluid-mechanic mechanism in a range of $SR = 0.5 - 1.0$.

It is particularly noticed that $Cl(t)$ for $SR = 0.5$ is larger than that for $SR = 1.0$ in the upstroke, and in other words the magnitude of the negative lift coefficient is deduced for $SR = 0.5$ after the effect of changing wing area is removed. There is the corresponding peak of $\Delta Cl(t)$ in the upstroke. To gain an insight into this phenomenon, a quasi-steady lifting line model is used in Appendix A to estimate the effects of the dynamic wing aspect ratio (AR) on the lift coefficient. According to the lifting line model, the magnitude of the negative lift coefficient in the upstroke is indeed decreased in the case of $SR = 0.5$ because the retracted wingspan in the upstroke induces the stronger downwash relative to the effective incoming flow. From this perspective, the lift enhancement in the upstroke is considerably contributed by the effects of the AR that reflects the change of the vortex strength induced by the downwash. However, the amplitude and phase of the lift coefficient predicted by the lifting line model are considerably different from those given by DNS since some relevant flow structures such as the leading-edge vortices in the unsteady separated flows are not taken into account.

**B. Effects of other parameters**

The time-averaged increment of the lift coefficient $\langle \Delta Cl \rangle_T$ is evaluated in the parametric subspace $(\phi, St, A^*, \alpha)$. Figure 6(a) shows $\langle \Delta Cl \rangle_T$ as a function of the phase angle $\phi$ for $SR = 0.5$, $St = 0.3$, $A^* = 0.25$ and $\alpha = 0^o$. There is the maximum in $\langle \Delta Cl \rangle_T$ at $\phi = 0.39\pi$ ($\phi = 70^o$). Figure 6(b) shows $\langle \Delta Cl \rangle_T$ as a function of the heaving amplitude $A^*$ for $SR = 0.5$, $\phi = \pi/2$, $St = 0.3$ and $\alpha = 0^o$, which indicates the maximum in $\langle \Delta Cl \rangle_T$ at $A^* = 0.375$. The dependency of $\langle \Delta Cl \rangle_T$ on the Strouhal number $St$ is shown in Fig. 6(c) for $SR = 0.5$, $\phi = \pi/2$, $A^* = 0.25$ and $\alpha = 0^o$, which indicates the maximum in $\langle \Delta Cl \rangle_T$ at $St = 0.4$. It is indicated that a preferred mode may exist in the subspace $(\phi, St, A^*, \alpha)$ to achieve the maximum of $\langle \Delta Cl \rangle_T$ although the true global optimal mode is not known yet until an optimization problem with suitable constraints is solved in the whole parametric space. Most importantly, the lift coefficient increment $\langle \Delta Cl \rangle_T$ in Fig. 6 is significantly larger than zero in all the cases. Therefore, the fluid-mechanic mechanism responsible to the lift enhancement is confirmed in the subspace $(\phi, St, A^*, \alpha)$ surveyed.

Figure 7 shows the time-averaged lift coefficient $<Cl>_T$ and increment $\langle \Delta Cl \rangle_T$ as a function of $\alpha$ for $SR = 0.5$, $\phi = \pi/2$, $St = 0.3$ and $A^* = 0.25$. For comparison, the time-averaged lift coefficient $<Cl_S>_T$ for the corresponding stationary wing with a fixed wingspan is plotted as a reference. The value of $\langle \Delta Cl \rangle_T$ is about $0.4 - 0.5$ in a range of $\alpha = 0 - 30°$ compared to the maximum of $<Cl_S>_T = 0.8$ for the stationary wing at $\alpha = 30°$. Therefore, it is again indicated that the enhancement of the lift coefficient by dynamically changing wingspan is significant.

As indicated Section II A, there is the formal relation $\langle \Delta Cl \rangle_T = C_{FS} \langle \Delta Cl_{S\_max} \rangle_T (a+b)/a$, where the correlation coefficient $C_{FS}$ is a function of the parameters $(SR, \phi, St, A^*, \alpha)$. Figure 8 shows the data of $\langle \Delta Cl \rangle_T$ and $\langle \Delta Cl_{S\_max} \rangle_T$ in all the cases, which indicates $0.3 \leq \langle \Delta Cl \rangle_T / \langle \Delta Cl_{S\_max} \rangle_T \leq 1$ depending on the parameters. This means that more than 30% of the overall lift enhancement is due to the change of the instantaneous lift coefficient $Cl$. In summary, the above parametric study indicates that the time-averaged increment of the lift coefficient $\langle \Delta Cl \rangle_T$ is significant which is mainly caused by the fluid-mechanic effect induced by dynamically changing the wingspan. The flow structures associated with this phenomenon will be explored in next section.

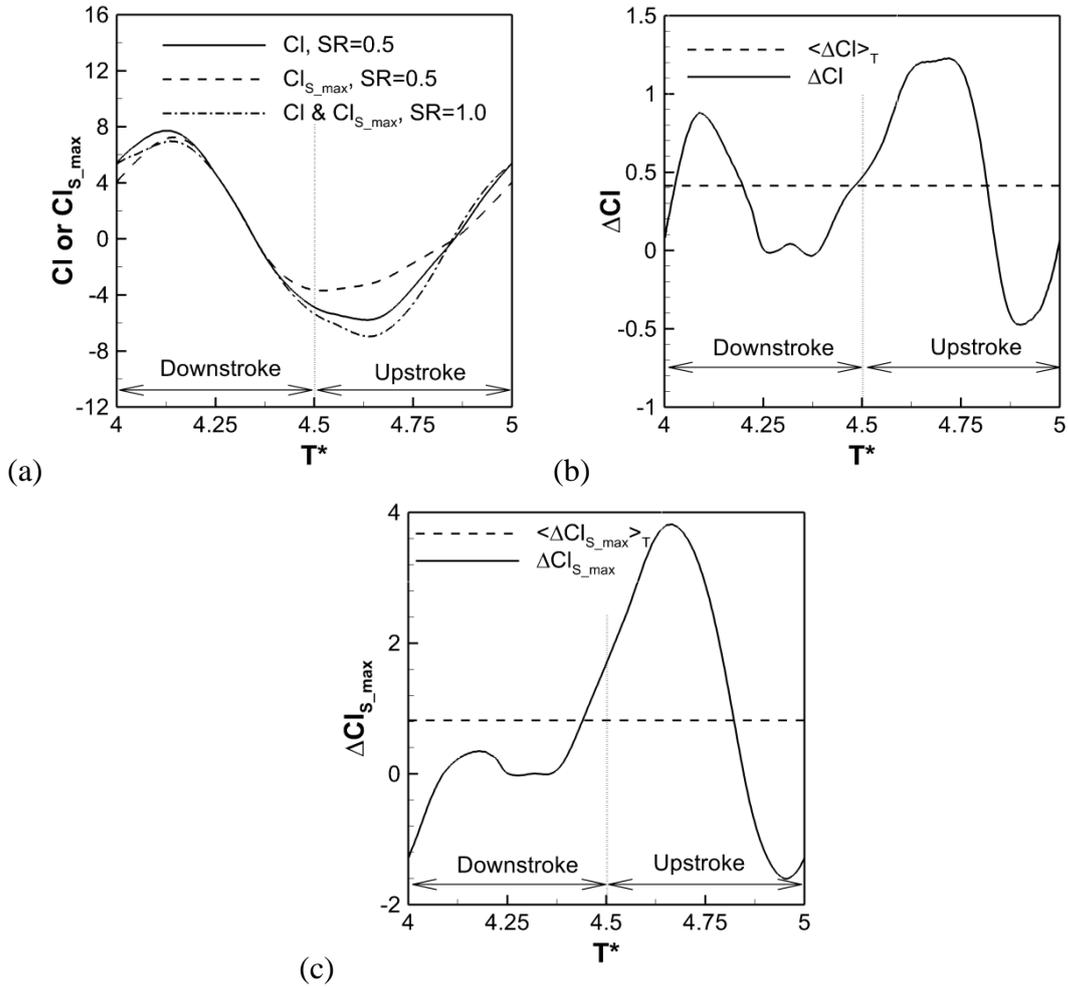

FIG. 4. The time histories of (a) $Cl$ and $Cl_{S\_max}$, (b) $\Delta Cl$, and (c) $\Delta Cl_{S\_max}$ in one period in the cases of $SR = 0.5$ and $SR = 1.0$ for $\phi = \pi/2$, $St = 0.3$, $A^* = 0.25$, and $\alpha = 0^o$. The time-averaged values $<\Delta Cl>_T$ and $<\Delta Cl_{S\_max}>_T$ are plotted for reference in (b) and (c), respectively.

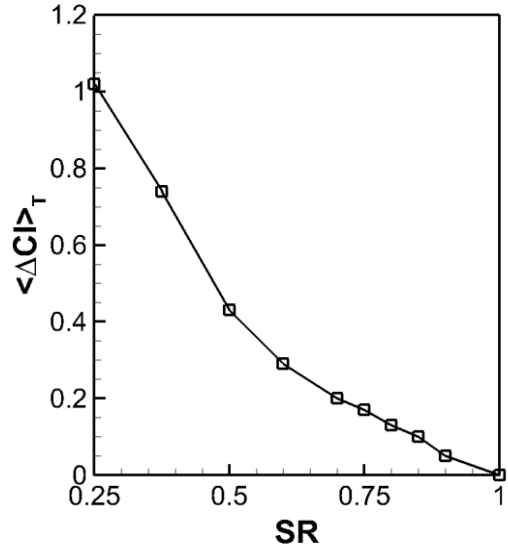

FIG. 5. The time-averaged increment of the lift coefficient $<\Delta Cl>_T$ as a function of the span ratio $SR$ for $\phi = \pi/2$, $St = 0.3$, and $A^* = 0.25$, and $\alpha = 0^o$.

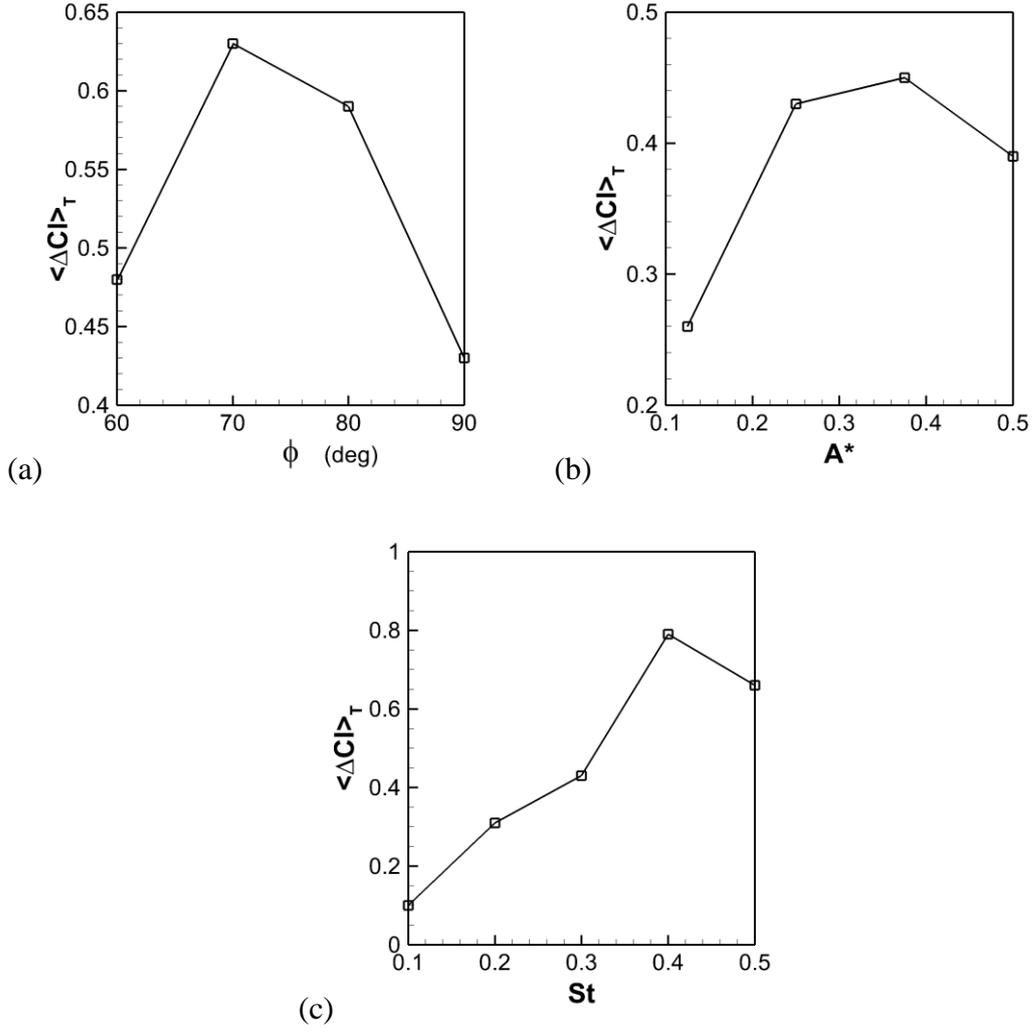

FIG. 6. The time-averaged increments of the lift coefficient $<\Delta Cl>_T$ as a function of (a) the phase difference for $SR=0.5$, $St=0.3$, $A^*=0.25$ and $\alpha=0^o$, (b) flapping amplitude for $SR=0.5$, $\phi=\pi/2$, $St=0.3$, and $\alpha=0^o$, and (c) Strouhal numbers for $SR=0.5$, $\phi=\pi/2$, $A^*=0.25$ and $\alpha=0^o$.

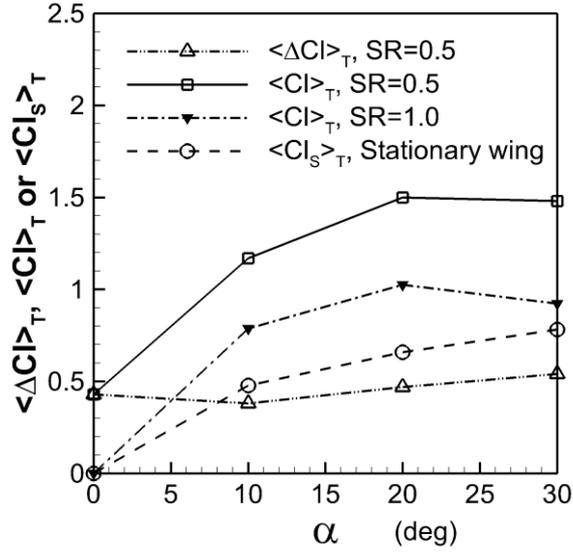

FIG. 7. The time-averaged lift coefficients $<Cl>_T$ and the increment $<\Delta Cl>_T$ at different AoAs for $\phi = \pi/2$, $St = 0.3$, and $A^* = 0.25$, where the time-averaged lift coefficient $<Cl_S>_T$ of the corresponding stationary wing are plotted as a reference.

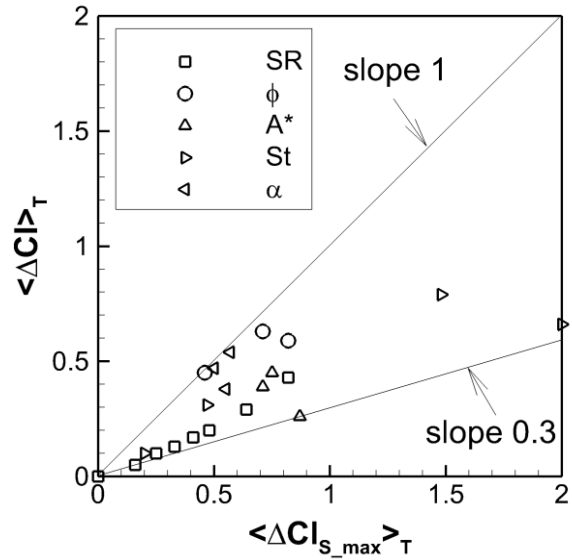

FIG. 8. The relation between $<\Delta Cl>_T$ and $<\Delta Cl_{S\_max}>_T$ in all the cases with the different parameters.

## IV. FLOW STRUCTURES

### A. General characteristics

To understand the physical mechanisms behind the lift enhancement associated with stretching and retracting wingspan, the flow structures in the cases of $SR = 1.0$ and $SR = 0.5$ for $(\phi, St, A^*, \alpha) = (\pi/2, 0.3, 0.25, 0^o)$ are investigated as the typical cases. The $Q$-criterion is used to identify the three-dimensional (3D) vortical structures, where $Q$ is the second invariant of the velocity gradient tensor. The flow structures in the case of $SR = 1.0$ are examined as a reference where the wingspan remains unchanged during the flapping. Then, the effects of stretching and retracting wingspan are studied by comparing the reference flow structures with those in the case of $SR = 0.5$. Figure 9 shows the typical instantaneous flow structures around the flapping rectangular wing in the cases of $SR = 1.0$ and $SR = 0.5$. In the case of $SR = 1.0$, the trailing-edge vortices (TEVs) and tip vortices (TVs) shedding from the wing are organized and connected to form vortex rings in the near wake. These vortex rings interact and braid into vortex chains as they travel downstream in the far wake. Two vortex rings are generated during each flapping cycle, as denoted by R1 and R2 in Fig. 9. In the case of $SR = 0.5$, the spanwise stretching and retracting during the flapping affects the flow structures around the wing, and a C-type vortex and vortex ring are generated during each period, which are denoted by C1 and R2 in Fig. 9, respectively.

The formation of the vortex rings in the near wake at five phases $T^* = 4.125$, $4.375$, $4.625$, $4.875$ and $5.125$ are shown in Fig. 10 in both the cases of $SR =$

*1.0* and *SR = 0.5*, where the top view of the flow structures are provided. In the case of *SR = 1.0*, as shown in Fig. 10(a), the LEV, TEV and TVs generated at the beginning of the downstroke are denoted by LEV1, TEV1, TV1 and TV2, respectively. The detached LEVs in the previous downstroke and upstroke are denoted by PLEV1 and PLEV2, respectively. As indicated in Fig. 10(b), LEV1, TV1 and TV2 are generated and attached to the wing, while TEV1 sheds from the trailing edge in the downstroke. LEV1 has the positive spanwise vorticity, contributing to the positive circulation around the wing and therefore the positive lift. Subsequently, new vortices LEV2, TEV2, TV3 and TV4 are generated during the reversal transition from the downstroke to upstroke, and LEV1, TV1 and TV2 generated during the downstroke shed from the wing surface. As shown in Fig. 10(c), TEV1, TEV2, TV1 and TV2 form a vortex ring R1. The detached vortex LEV1 stays near the upper surface of the wing and travels to the downstream, as shown in Figs. 10(c)-(e). The vortex PLEV1 generated in the previous downstroke mergers with TEV2 newly generated during the reversal transition from downstroke to upstroke, as shown in the vorticity contours in Fig. 11(c).

In the case of *SR = 0.5*, the evolution of the flow structures is similar. At the beginning of the downstroke, as shown in Fig. 10(a), the vortices LEV1, TEV1, TV1 and TV2 are generated. The detached leading-edge vortices in the previous downstroke and upstroke are denoted by PLEV1 and PLEV2. The leading-edge vortex LEV1 is generated and attached to the leading edge during the downstroke. The trailing-edge vortex TEV1 sheds from the trailing edge shortly after the beginning of downstroke.  The tip vortices TV1 and TV2 start to shed at the middle of the downstoke due to the retraction of the wingspan in this case, which is different from the case of *SR = 1.0* where TV1 and TV2 shed at the end of the downstroke.  At the reversal transition from downstroke to upstroke, as shown in Fig. 10(c), LEV1 sheds from the leading edge, and then TEV1, LEV1, TV1 and TV2 forms the vortex ring R1 during this reversal.  It is noticed that TV1 and TV2 are not connected to the new generated TEV2, which is different from the case of *SR = 1.0*.  After shedding from the leading edge, LEV1 stays near the upper surface of the wing, contributing the positive lift.  Therefore, the vortex capture mechanism also exists during the downstroke in this case of *SR = 0.5*.  The vortex PLEV1 originated from the previous downstroke merges with the new generated TEV2 during this reversal, as shown in the vorticity contours in Fig. 11(c).

The contributions of the leading-edge vortices on the upper and lower surfaces to the lift can be directly evaluated by using the general lift formula given in Section V.  Nevertheless, as shown in Figs. 10 and 13, the lift generation can be inferred from the wake structures as the footprints of the vortical structures developed from the wing.  In the case of *SR = 0.5*, the vortex ring in the wake generated in the downstroke is much larger than that generated in the upstroke.  In contrast, in the case of *SR = 1.0*, the vortex rings generated in the downstroke is similar in the size to that in the upstroke.  As a result, the total momentum induced by the larger ring in the downstroke has the larger magnitude than that in the upstroke in the case of *SR = 0.5*, leading to the higher lift.

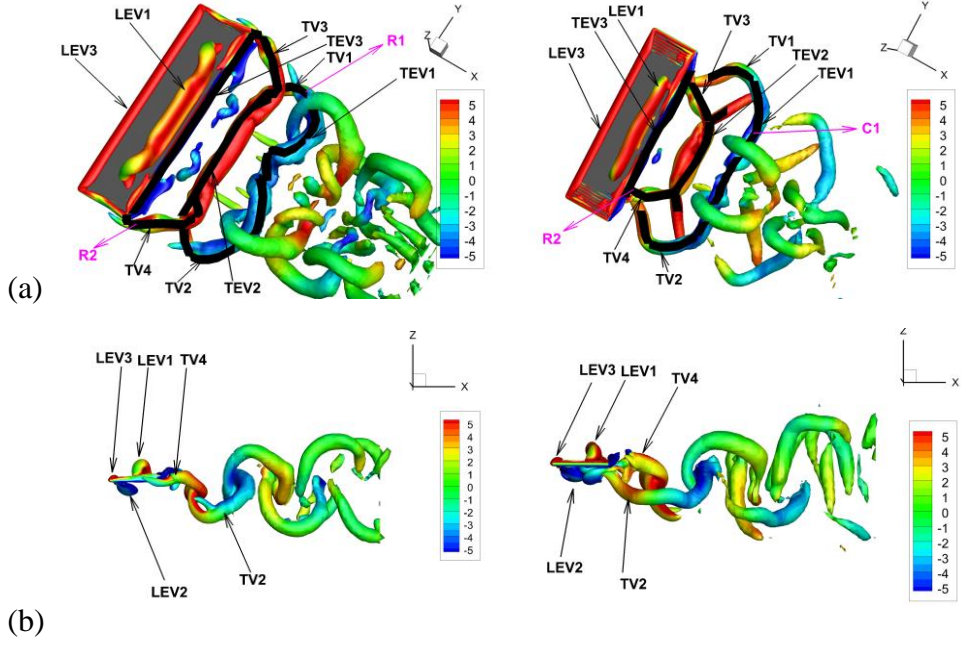

FIG. 9. The typical instantaneous flow structures around the flapping rectangular wing in the cases of $SR = 1.0$ (right) and $SR = 0.5$ (left): (a) perspective view with $Q = 3.0$, and (b) side view with $Q = 3.0$. The vortical flow structures are identified using the $Q$-criterion. The pseudo-color on the flow structures shows the values of the spanwise voriticity. The vortex rings are highlighted using the solid black line. The detailed parameters for the cases of $SR = 1.0$ and $SR = 0.5$ are $(SR, \phi, St, A^*, \alpha) = (1.0, \pi/2, 0.3, 0.25, 0^o)$ and $(SR, \phi, St, A^*, \alpha) = (0.5, \pi/2, 0.3, 0.25, 0^o)$, respectively.

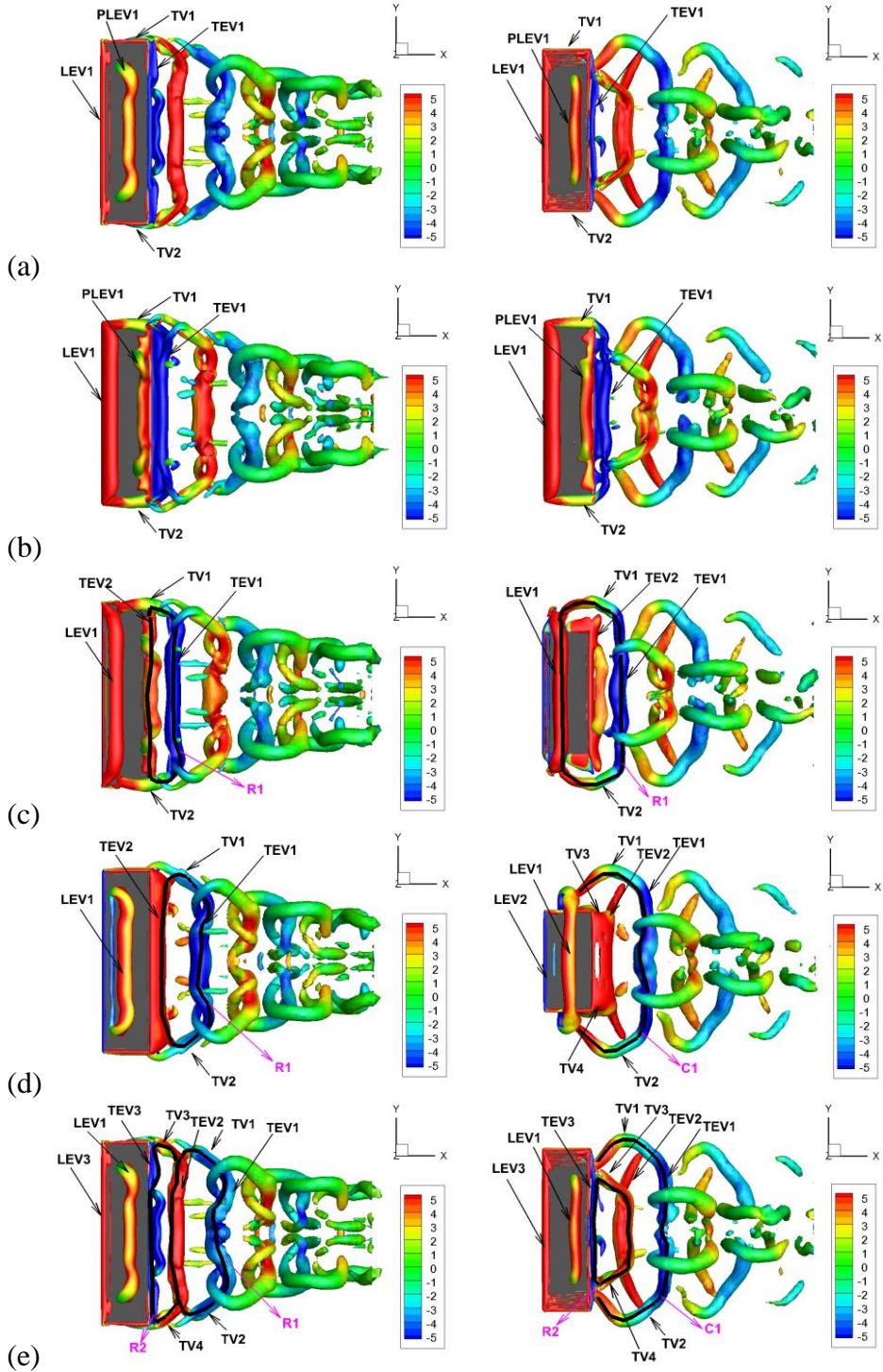

FIG. 10. The three-dimensional flow structures viewed from the top in the cases of *SR = 1.0* (left) and *SR = 0.5* (right) at different time (a) $T^* = 4.125$, (b) $T^* = 4.375$, (c) $T^* = 4.625$, (d) $T^* = 4.875$, and (e) $T^* = 5.125$. The iso-surface of $Q = 3.0$ is shown, where the colors indicate the spanwise vorticity. The detailed parameters for the cases of *SR = 1.0* and *SR = 0.5* are $(SR, \phi, St, A^*, \alpha) = (1.0, \pi/2, 0.3, 0.25, 0^o)$ and $(SR, \phi, St, A^*, \alpha) = (0.5, \pi/2, 0.3, 0.25, 0^o)$, respectively.

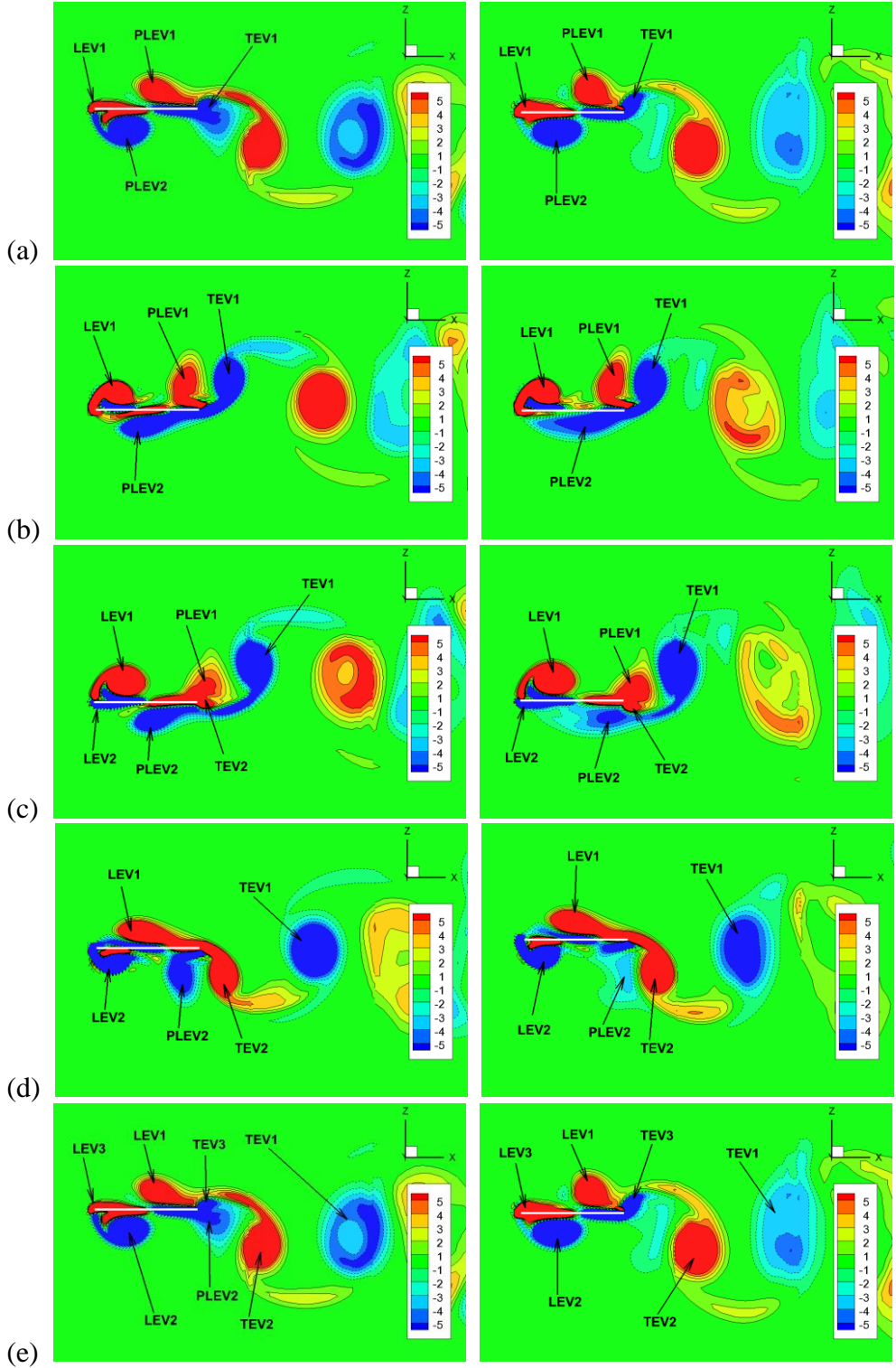

FIG. 11. The contours of the spanwise vorticity in $y = 0$ in the cases of $SR = 1.0$ (left) and $SR = 0.5$ (right) at different time (a) $T^* = 4.125$, (b) $T^* = 4.375$, (c) $T^* = 4.625$, (d) $T^* = 4.875$, and (e) $T^* = 5.125$. The detailed parameters for the cases of $SR = 1.0$ and $SR = 0.5$ are $(SR, \phi, St, A^*, \alpha) = (1.0, \pi/2, 0.3, 0.25, 0^o)$ and $(SR, \phi, St, A^*, \alpha) = (0.5, \pi/2, 0.3, 0.25, 0^o)$, respectively.

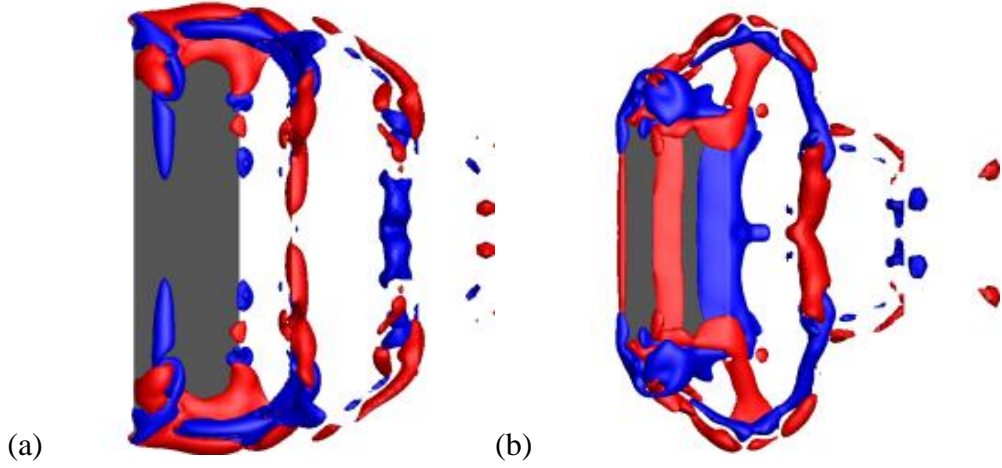

FIG. 12. The top-viewed iso-surfaces of the spanwise vortex stretching term $\boldsymbol{\omega}\cdot\nabla u_y$ in the cases of (a) *SR = 1.0* and (b) *SR = 0.5* at $T^*=4.625$ in the upstroke for $\phi=\pi/2$, $St=0.5$, $A^*=0.25$ and $\alpha=0°$. The red color shows the iso-surfaces of $\boldsymbol{\omega}\cdot\nabla u_y = 10$, indicating the regions where the vortices are stretched. The blue color shows the iso-surfaces of $\boldsymbol{\omega}\cdot\nabla u_y = -10$, indicating the regions where the vortices are compressed.

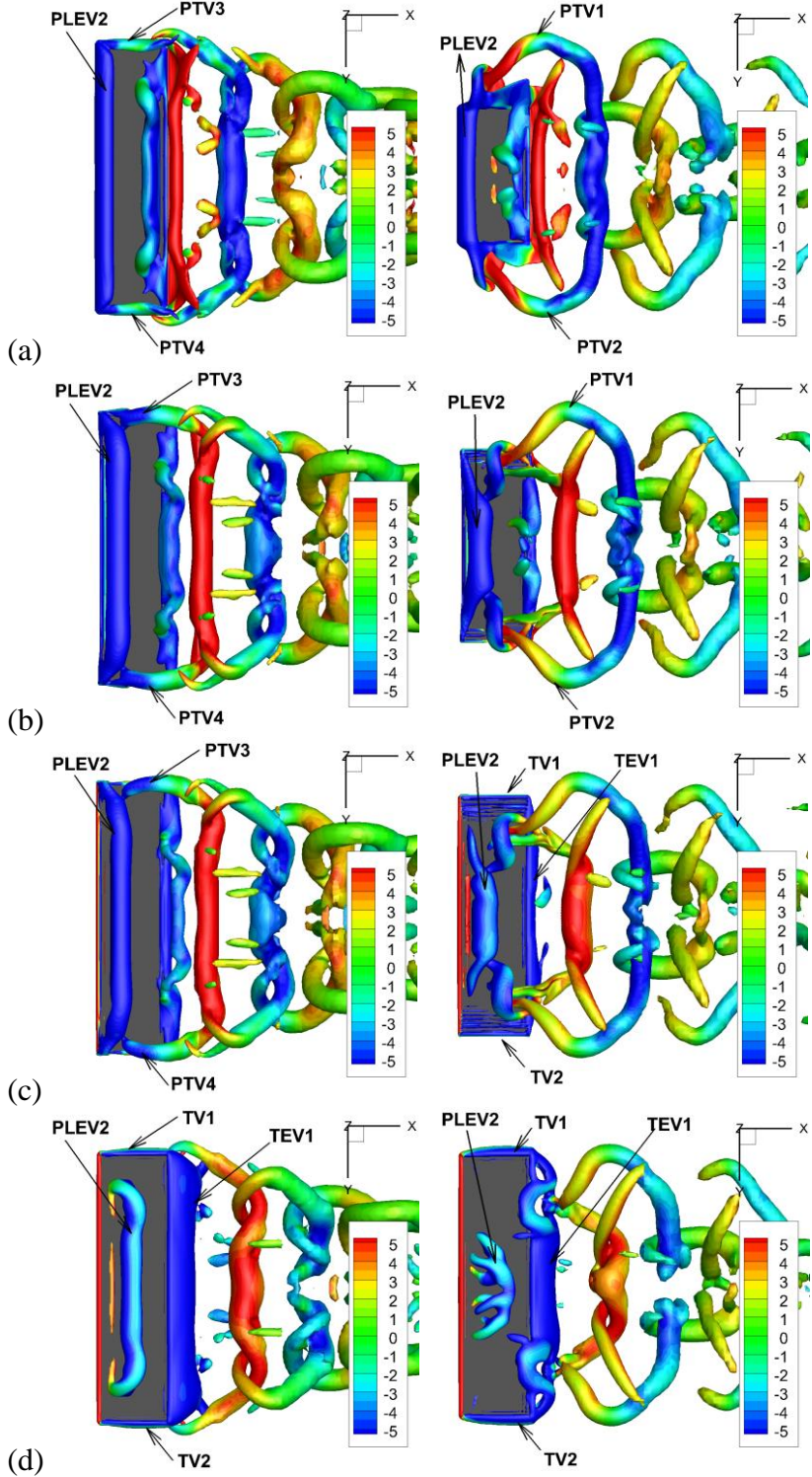

FIG. 13. The development of LEVs on the lower surface viewed from the bottom at (a) $T^* = 3.75$ and (b) $T^* = 4.0$ in the upstroke, and (c) $T^* = 4.125$ and (d) $T^* = 4.375$ in the downstroke. The left column: $SR = 1.0$, and the right column: $SR = 0.5$. The iso-surfaces of $Q = 3.0$ are shown, and the colors indicate the spanwise vorticity. The detailed parameters for the cases of $SR = 1.0$ and $SR = 0.5$ are $(SR, \phi, St, A^*, \alpha) = (1.0, \pi/2, 0.3, 0.25, 0^o)$ and $(SR, \phi, St, A^*, \alpha) = (0.5, \pi/2, 0.3, 0.25, 0^o)$, respectively.

## B. Vortex capture and stretching

Figure 11 shows the contours of the spanwise vorticity in the symmetrical plane $y = 0$ in the cases of $SR = 1.0$ and $SR = 0.5$ at the phases $T^* = 4.125$, $T^* = 4.375$, $T^* = 4.625$, $T^* = 4.875$, and $T^* = 5.125$. In both the cases, the vortex LEV1 is generated, and the vortex PLEV1 is trapped near the upper surface of the wing for about $1.5$ periods, which contributes to the generation of the positive lift until it merges with TEV2. This is referred to as the vortex capture mechanism in flapping flight, which is similar to the capture of a free vortex on an airfoil [44-46]. In two dimensions, although a free vortex cannot be stabilized near a stationary airfoil, vortex capture seems feasible for lift enhancement on a moving wing with the right kinematics in a certain period. This phenomenon also occurs in hovering flight of insects [7, 9, 10, 47]. To compare the vortex stretching mechanisms in the cases of $SR = 1.0$ and $SR = 0.5$, the top-viewed iso-surfaces of the spanwise vortex stretching term $\omega \cdot \nabla u_y$ at $T^* = 4.625$ in the upstroke are shown in Fig. 12 for $\phi = \pi/2$, $St = 0.5$, $A^* = 0.25$ and $\alpha = 0°$. In the case of $SR = 0.5$, it is found that the vortex LEV1 near the upper surface is consistently intensified by spanwise vortex stretching, which is supported by the correlations between the time histories of the integrated spanwise vortex stretching, vorticity and wing motion shown in Fig. 15(a). The intensified vortex LEV1 contributes to the larger positive lift even though it is shorter, which is further confirmed by the calculation of the vortex lift in the upstroke in Section V.

On the lower surface, the vortex PLEV2 generated in the previous cycle stays for about 1.5 periods near the lower surface of the wing. However, PLEV2 has the negative spanwise vorticity, which contributes to the negative lift as the negative vortex capture. More interestingly, as shown in Figs. 11(c)-(e), the vortex PLEV2 originated from the previous upstroke on the lower surface in the case of *SR = 0.5* is much weaker than that in the case of *SR = 1.0*. This weaker PLEV2 is related to the spanwise stretching and retracting wingspan. To observe the detailed 3D structure of PLEV2, the development of PLEV2 on the lower surface in the case of *SR = 0.5* is shown in Fig. 13 in comparison with that in the case of *SR = 1.0* as a reference. PLEV2 generates at $T^* = 3.5$, and part of PLEV2 sheds during $T^* = 3.5 \sim 4.0$ in the upstroke due to the retracting of wingspan. In the upstroke, PLEV2 for *SR = 0.5* is much shorter than that for *SR = 1.0*. Therefore, the total contribution of PLEV2 to the negative lift in the case of *SR = 0.5* is smaller than that in the case of *SR = 1.0*. As the wing moves downward after $T^* = 4.0$, PLEV2 sheds from the lower surface of the wing with several legs, and it deforms into streamwise stripes downstream (see Fig.13(c)). Due to this redistribution of vorticity, the spanwise vorticity of PLEV2 is smaller in the downstroke. Figure 14 shows the bottom-viewed iso-surfaces of the spanwise vortex stretching term $\boldsymbol{\omega} \cdot \nabla u_y$ in the cases of *SR = 1.0* and *SR = 0.5* at $T^* = 4.0$ (the end of the upstroke) for $St = 0.5$, $\phi = \pi/2$, $A^* = 0.25$ and $\alpha = 0°$. The spanwise vortex stretching on the lower surface in the case of *SR = 0.5* is highly 3D, which corresponds to the complicated 3D vortical structures observed in Fig. 13.

To quantitatively evaluate the correlations between the spanwise vortex stretching and wingspan motion in a flapping period, the two volume integrals of the spanwise vortex stretching term on the upper and lower surfaces around the LEVs are evaluated, which are defined as

$$[S_y]_{upper} \approx \int_{D_{upper}} \omega_y^* \partial u_y^* / \partial y^* \, dV^*, \quad [S_y]_{lower} \approx \int_{D_{lower}} \omega_y^* \partial u_y^* / \partial y^* \, dV^*, \qquad (14)$$

where $D_{upper}$ and $D_{lower}$ are the selected domains $[-5,0.5] \times [-3,3] \times [z_w^*, 14]$ and $[-5,0.5] \times [-3,3] \times [-14, z_w^*]$ on the upper and lower surfaces, respectively, to include all the spanwise vorticities near the LEVs and exclude the vorticities in the wake. Here the superscript * denotes the non-dimensional quantities. Similarly, the strength of the spanwise vorticity in the two domain are given by $[\omega_y]_{upper}$ and $[\omega_y]_{lower}$. The spanwise motion of the wing is characterized by the spanwise velocity $v_M$ of a middle point on the surface between the wing root and tip. Figure 15 shows the time histories of $[\omega_y]$, $[S_y]$ and $v_M$ in a flapping period in the case of $SR = 0.5$. On the upper surface, as indicated in Fig. 15(a), the integrated spanwise vortex stretching is closely correlated to the spanwise motion of the wing, and they are also in phase. The integrated spanwise vorticity has a phase shift of about $100^\circ$ relative to the integrated spanwise vortex stretching. The phase shift is expected since the time derivative of the volume-integrated vorticity is proportional to the volume-integrated vortex stretching term in the integral form of the vorticity equation in the domain, i.e., $d[\omega_y]/dt \approx [S_y] + [B]$, where $[B]$ is the contribution from the outer control surface. In contrast, on the lower surface, there is a phase shift of $180^\circ$ between $[S_y]$ and $v_M$ as shown in Fig. 15(b). The magnitude of the volume-integrated spanwise vorticity on the upper surface is larger than that on the lower surface in the most of the downstroke and the first third of the upstroke, which is consistent with the vortex lift enhancement discussed in Section V.

Based on the above observations, the leading-edge vortices on the upper surface are significantly intensified by spanwise vortex stretching associated with dynamically changing wingspan, which contributes to the increased lift in the case of of $SR = 0.5$. At the meantime, the shorter and weaker leading-edge vortices on the lower surface in the upstroke have the smaller contribution to the negative lift. The combination of these mechanisms results in the overall vortex lift enhancement that corresponds to the peak in $\Delta Cl$ in Fig. 4 in the upstroke in the case of $SR = 0.5$. In the downstroke, since the vortical structures on the lower surface become highly 3D, the spanwise vorticity of these structures is decreased. As a result, they have the smaller contribution to the negative lift, which leads to the smaller peak in $\Delta Cl$ in Fig. 4 in the downstroke.

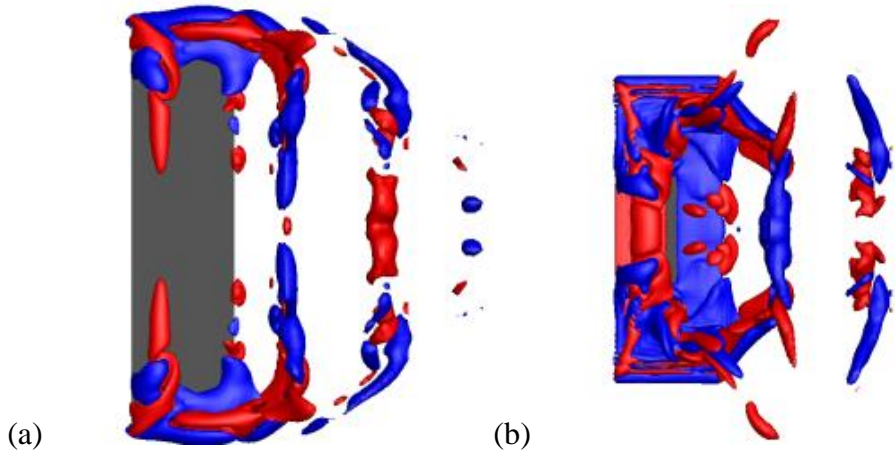

(a)          (b)

FIG. 14. The bottom-viewed iso-surfaces of the spanwise vortex stretching term $\omega \cdot \nabla u_y$ in the cases of (a) $SR = 1.0$ and (b) $SR = 0.5$ at $T^* = 4.0$ (the end of the upstroke) for $\phi = \pi/2$, $St = 0.5$, $A^* = 0.25$ and $\alpha = 0°$. The red color shows the iso-surfaces of $\omega \cdot \nabla u_y = 10$, indicating the regions where the vortices are stretched. The blue color shows the iso-surfaces of $\omega \cdot \nabla u_y = -10$, indicating the regions where the vortices are compressed.

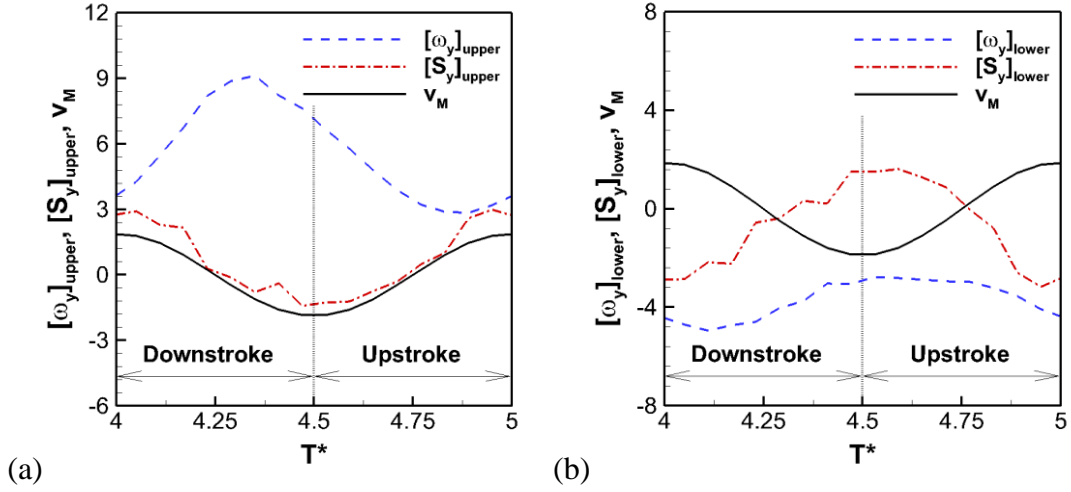

(a) (b)

FIG. 15. The time histories of the integrated vortex stretching, vorticity and spanwise motion velocity in a flapping period on (a) the upper surface and (b) the lower surface in the case of $SR = 0.5$.

## V. DECOMPOSITION OF LIFT: VORTEX FORCE AND LOCAL ACCELERATION

The lift can be decomposed to further understand the relationship between the lift enhancement and the flow structures. The force acting on a body immersed in a fluid flow is usually calculated by

$$\boldsymbol{F} = \oint_{\partial B} (-p\boldsymbol{n} + \boldsymbol{\tau})dS, \qquad (15)$$

where, $p$ and $\boldsymbol{\tau}$ are the pressure and skin friction on the surface of a body, respectively, $\boldsymbol{n}$ is the normal direction of the body surface $\partial B$ pointing to the outside of the body.

By using the Navier-Stokes equations and Gauss's theorem, Eq. (15) can be expressed as

$$\boldsymbol{F} = \rho \int_{V_f} \boldsymbol{u} \times \boldsymbol{\omega}\, dV - \rho \int_{V_f} \frac{\partial \boldsymbol{u}}{\partial t}\, dV - \oint_{\Sigma} \left(p + \rho q^2/2\right) \boldsymbol{n}\, dS + \oint_{\Sigma} \boldsymbol{\tau}\, dS - \rho \oint_{\partial B} \left(q^2/2\right) \boldsymbol{n}\, dS$$

, (16)

where $\boldsymbol{u}$ is the velocity, $\boldsymbol{\omega}$ is the vorticity, $q = |\boldsymbol{u}|$, $\partial B$ denotes a solid boundary of the body domain $B$, $V_f$ denotes the control volume of fluid, $\Sigma$ denotes an outer control surface in which the body is enclosed, and $\boldsymbol{n}$ is the unit normal vector pointing to the outside of a control surface. The first term in the right-hand side (RHS) of Eq. (16) is a volume integral of the Lamb vector $\boldsymbol{l} = \boldsymbol{u} \times \boldsymbol{\omega}$ that represents the vortex force. The second term is a volume integral of the local acceleration of fluid induced by a moving solid body for the unsteady inertial effect. The third and fourth terms are the surface integrals of the total pressure $p + \rho q^2/2$ and the surface shear stress on the outer control surface $\Sigma$. The fifth term is the boundary term. In an inviscid irrotational unsteady flow where the first, third and fourth terms in Eq. (16) vanish, the remaining second and fifth terms together are interpreted as the added mass force in ideal fluid mechanics. The fifth term is interpreted as the part of the added mass force associated with the fluid virtually occupying the body domain $B$ (a virtual fluid body).

The lift on a body is given by $F_z = \boldsymbol{k} \cdot \boldsymbol{F}$, where $\boldsymbol{k}$ is the unit vector normal to the freestream,. A rectangular domain $D$ is selected as a control volume to simplify the lift expression. For a sufficiently large rectangular outer control surface $\Sigma$, as shown in Fig. 16, computations indicates that the third, fourth and fifth terms in the RHS of Eq. (16) are small. In the typical cases considered in this work, the relative error in lift estimation caused by neglecting these terms is less than 3%.

Therefore, the main terms in the RHS of Eq. (16) are the first and second terms. The first and second terms represent the vortex lift acting on the wing (the Lamb vector integral projected onto the normal direction to the freestream) and the effect of local fluid acceleration, respectively.

To investigate the roles of the flow structures in the lift generation, a rectangular control volume of $[-5, 0.5] \times [-3, 3] \times [-14, 14]$ in the streamwise, spanwise and vertical directions is selected. The bottom and top boundaries of the control volume are located at $z = \pm 14$. Figure 16 shows the lift coefficient $Cl_{simp}$ calculated based on the simplified lift formula with only the vortex lift and the local acceleration terms [see Eq. (16)] that are denoted by $Cl_{vort}$ and $Cl_{acc}$, respectively. As shown in Fig. 16, the lift coefficient $Cl_{simp}$ is in good agreement with $Cl$ calculated based on the pressure and viscous stress fields on the wing. In the case of *SR = 0.5*, the time-averaged lift coefficient calculated by using the simple lift formula is 0.43, which agrees with 0.42 given by calculation based on the pressure and viscous stress fields on the wing surface. The relative error is about 2.4%. For the case of *SR = 1.0*, the positive lift and negative lift generated in the flapping are canceled out each other due to the symmetrical flapping motion such that both the time-averaged lift coefficients $Cl$ and $Cl_{simp}$ are zero.

Figure 17 shows the contributions of the Lamb vector term or vortex force ($\Delta Cl_{vort}$) and local acceleration term ($\Delta Cl_{acc}$) to the increment of the lift coefficient $\Delta Cl$ in one period. It is indicated that the contribution of the vortex force to $\Delta Cl$ is positive in a full period particularly in the upstroke. This means that the vortex force is enhanced by stretching and retracting wingspan during the flapping. In this case, the local acceleration term has the negative contribution to $\Delta Cl$ particularly in the upstroke. The difference of the local acceleration term of the lift coefficient, $\Delta Cl_{acc}$, is related to the time derivative of the vortex-induced velocity that is approximately proportional to the time derivative of vorticity and therefore the vortex stretching term associated with dynamically changing wingspan. It is indeed found that the time histories of $\Delta [S_y]$ and $\Delta Cl_{acc}$ are correlated, where $\Delta [S_y] = [S_y](SR=0.5) - [S_y](SR=1)$ is the difference of the volume-integrated spanwise vortex stretching, $[S_y] = [S_y]_{upper} + [S_y]_{lower}$ is the volume-integrated spanwise vortex stretching in the whole domain, and $[S_y]_{upper}$ and $[S_y]_{lower}$ are defined in Eq. (14).

Further, the contributions of the vortical structures to the lift coefficient $Cl$ in the upper and lower portions of the control volume divided by the flat-plate rectangular wing are evaluated in the cases of $SR = 0.5$ and $SR = 1.0$. Figure 18 shows the contributions of the Lamb vector term to $Cl$ in the upper and lower portions of the control volume in one period. In the average sense, the contributions of the Lamb vector term in the upper and lower portions to $Cl$ are positive and negative, respectively, in the cases of $SR = 0.5$ and $SR = 1.0$. During the upstroke, the contributions of the vortex force to $Cl$ in both the upper and lower portions for $SR = 0.5$ are larger than those for $SR = 1.0$. As pointed out in Section IV B, the leading-edge vortices on the upper surface in the upstroke are significantly intensified by spanwise vortex stretching, which contributes to the increased lift in the case of of *SR = 0.5*. At the meantime, the shorter and more 3D leading-edge vortices on the lower surface have the smaller contribution to the negative lift. These differences lead to the higher positive peak in *ΔCl(t)* during the upstroke in Fig. 4. During the downstroke, the contribution of the vortex force to *Cl* in the lower portion for *SR = 0.5* is still larger than that for *SR = 1.0* although the contributions in the upper portion in both the cases remain almost the same. As a result, there is the smaller peak in *ΔCl(t)* during the downstroke as indicated in Fig. 4. This is related to the vortical structures with the decreased spanwise vorticity on the lower surface discussed in Section IV B. In summary, the vortex force associated with the vortical structures altered by dynamically changing wingspan significantly contributes the lift enhancement.

In addition, to visualize the contribution of the vortical structures to the lift, the instantaneous contours of the Lamb vector projected in the vertical direction at *T\* = 4.875* (when the wing is near the end of the upstroke) are shown in Fig. 19. The positive vertical component of the Lamb vector associated with the intensified vortical structures on the upper surface is increased in the case of *SR = 0.5*. On the lower surface, as shown in in Fig. 19(b), the negative vertical component of the Lamb vector in the case of *SR = 0.5* has a smaller magnitude than that in the case of *SR = 1.0*. This corresponds to the vortex PLEV2 that has much smaller spanwise vorticity, as shown in Fig. 13(d). Therefore, the captured vortical structures that are responsible to the vortex lift are altered by stretching and retracting wingspan during the flapping, which contributes the increased time-averaged lift. The induced drag associated with the enhanced lift is discussed in Appendix B.

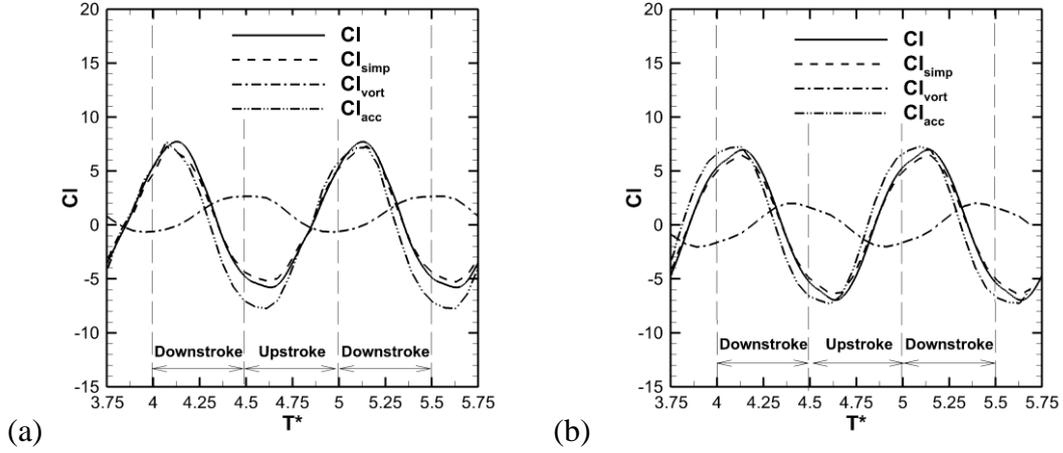

FIG. 16. The decomposition of the lift coefficient in (a) *SR = 0.5*, and (b) *SR = 1.0* for $\phi = \pi/2$, $St = 0.3$, $A^* = 0.25$ and $\alpha = 0^o$.

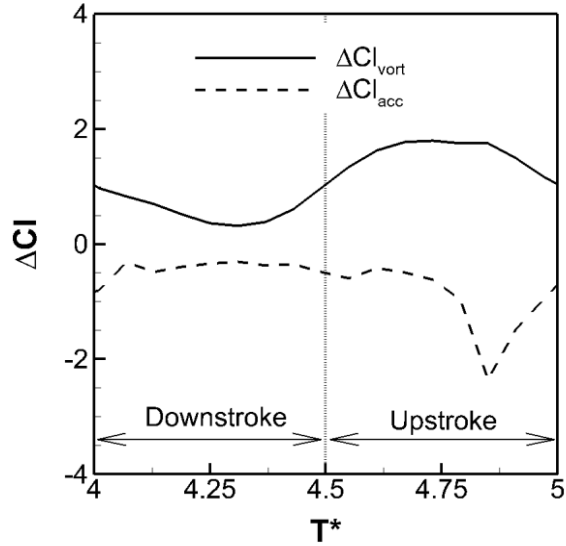

FIG. 17. The contributions of the Lamb vector term (the vortex force) and local acceleration term to the increment of the lift coefficient *ΔCl* in one period for $\phi = \pi/2$, $St = 0.3$, $A^* = 0.25$ and $\alpha = 0^o$.

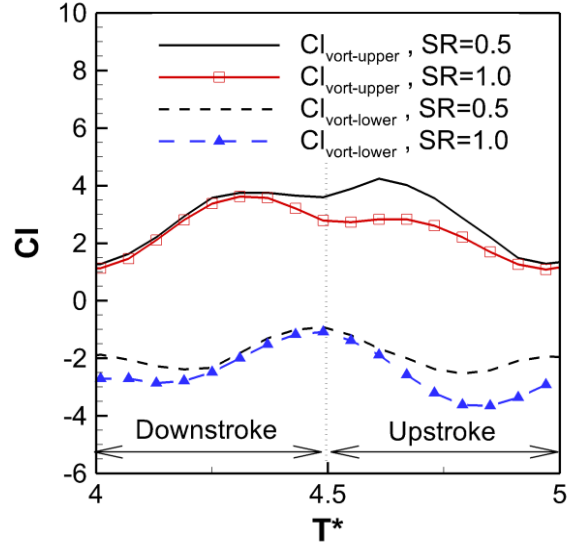

FIG. 18. The contributions of the Lamb vector term to the lift coefficient $Cl$ in the upper and lower portions of the control volume for $\phi = \pi/2$, $St = 0.3$, $A^* = 0.25$ and $\alpha = 0^o$.

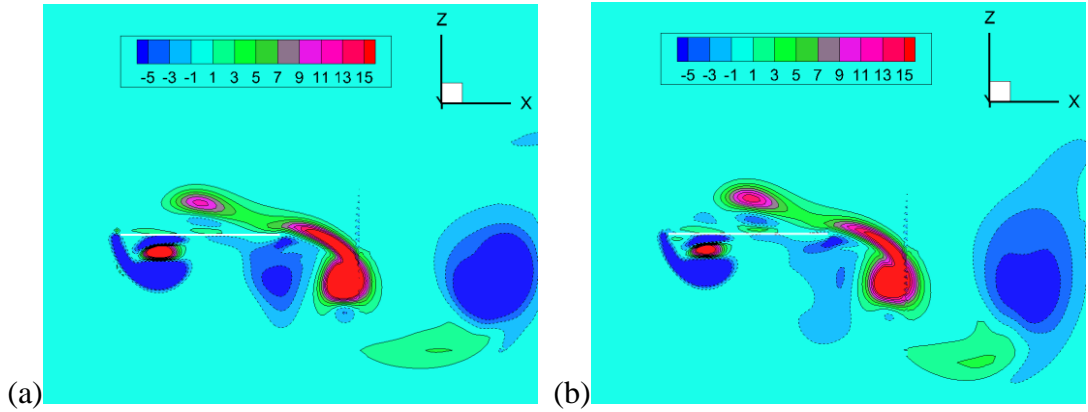

(a) (b)

FIG. 19. The vertical component of the Lamb vector in the symmetrical plane $y = 0$ at $T^* = 4.875$ in the upstroke for (a) $SR = 1.0$ and (b) $SR = 0.5$ for $\phi = \pi/2$, $St = 0.3$, $A^* = 0.25$ and $\alpha = 0^o$.

## VI. CONCLUSIONS

    The flapping flat-plate rectangular wing with a dynamically stretching and retracting wingspan is studied through direct numerical simulations as a model of biologically-inspired morphing wing for lift enhancement. The wingspan varies in

the given kinematics in a flapping cycle. The detailed flow structures and unsteady lift of the wing at the Reynolds number of 300 are obtained. The calculations in the parametric space consisting of the span ratio, phase angle, Strouhal number, heaving amplitude and geometrical angle of attack indicate that the lift coefficient is significantly enhanced by stretching and retracting wingspan during the flapping flight. The lift enhancement is achieved by not only the effect of changing the wing area, but also the fluid-mechanic mechanism induced by dynamically changing wingspan. The fluid-mechanic mechanism is further confirmed by the observation that the leading-edge vortices on the upper surface in the upstroke are significantly intensified by spanwise vortex stretching associated with dynamically changing wingspan, contributing to the increased lift. At the meantime, the shorter and weaker leading-edge vortices on the lower surface in both the upstroke and downstroke have the smaller contribution to the negative lift. These mechanisms lead to the overall vortex lift enhancement in the upstroke and downstroke, and the increased time-averaged lift coefficient. Furthermore, the simple lift decomposition indicates that the lift enhancement results mainly from the increase of the vortex lift induced by dynamically changing wingspan in flapping flight. This work reveals the significance of the dynamic wing morphing in the lift generation that is ubiquitous in bird and bat flight. The further implication is that birds and bats could manipulate the spanwise wing stretching-and-retracting motion in flapping flight unlike insects that do not change their wingspan. This may illustrate a fundamentally different aerodynamic aspect between birds/bats and insects in flight.

**ACKNOWLEDGEMENTS**

This work was supported by the National Natural Science Foundation of China under Project Nos. 10872201, 11232011, 11302238 and 11372331, and the National Basic Research Program of China (973 Program) under Project No. 2013CB834100 (Nonlinear science). Tianshu Liu would like to acknowledge the hospitality received at LNM during his visit where he accomplished this work. The simulations were performed on TianHe-1. The authors would like to acknowledge the support from National Supercomputer Center in Tianjin.**ACKNOWLEDGEMENTS**

This work was supported by the National Natural Science Foundation of China under Project Nos. 10872201, 11232011, 11302238 and 11372331, and the National Basic Research Program of China (973 Program) under Project No. 2013CB834100 (Nonlinear science). Tianshu Liu would like to acknowledge the hospitality received at LNM during his visit where he accomplished this work. The simulations were performed on TianHe-1. The authors would like to acknowledge the support from National Supercomputer Center in Tianjin.

**APPENDIX A. QUASI-STEADY LIFTING LINE MODEL**

To provide an understanding into the effects of the dynamic wing aspect ratio (AR) on the lift coefficient, a quasi-steady lifting line model is used in which the vertical position of the lifting line is described by Eq. (3) and the dynamic AR is described by Eq. (4). It is assumed that the trailing wake vortex sheet moves with the lifting line vertically like a flat rigid plate. By applying the classical lifting line theory as a quasi-steady model [48], the instantaneous lift coefficient is given by

$$Cl(t) = \frac{a_0 \alpha_{eff}}{1+(a_0/\pi AR)(1+\tau)}, \quad (A1)$$

where $a_0 = 2\pi$ is the lift slope of the 2D airfoil, the effective AoA is approximately given by $\alpha_{eff} = tan^{-1}(-\dot{z}_w/U_\infty) + \alpha_0 = tan^{-1}[-\pi St \cos(2\pi f t)] + \alpha_0$, the dynamic AR is $AR = AR_0[a - b\sin(2\pi f t + \phi)]$, the vertical velocity of the lifting line is $\dot{z}_w = 2\pi f A \cos(2\pi f t)$, the flapping Strouhal number is $St = 2fA/U_\infty$, and $\alpha_0 = \alpha_g + \alpha_{L=0}$ is the sum of the geometrical AoA and zero-lift AoA associated with the wing camber distribution. The coefficient $\tau$ is related to the wing planform, which is time-dependent in this case (typically $\tau = 0 - 0.25$). In the first-order approximation for a preliminary estimation, we simply set $\tau = 0$ that corresponds to the optimum planform (e.g. the elliptical planform).

Figure A1 shows the instantaneous lift coefficients given by Eq. (A1) as a function of the non-dimensional time $T^* = ft - 1/4$ in the cases of $SR = 0.5$ and $SR = 1.0$ (the fixed wingspan), where the DNS results are plotted for comparison. In both the cases, $(\phi, St, A^*, \alpha_0) = (\pi/2, 0.3, 0.25, 0^o)$. The lifting line model indicates that the magnitude of the negative lift coefficient in the upstroke is decreased in the case of *SR = 0.5* because the retracted wingspan in the upstroke induces the stronger downwash relative to the effective incoming flow.   This change in the upstroke is consistent with the DNS results although the amplitude and phase of the lift coefficient predicted by the quasi-steady lifting line model are considerably different.   Figure A2 shows the instantaneous increment of the lift coefficient $\Delta Cl(t) = Cl(t) - Cl(t; SR=1)$ as a function of $T^*$, where the DNS result is plotted for comparison.   Interestingly, as shown in Fig. A2, such a simple quasi-steady model shows the significant lift enhancement in the upstroke, which is qualitatively consistent with the DNS result.   In this sense, the effects of the dynamic AR significantly contribute the peak of $\Delta Cl(t)$ in the upstroke.   At the same time, the induced drag is increased as indicated in Fig. 20.   From a standpoint of the thin-airfoil theory, the effects of the AR in the upstroke essentially reflect the weakened vortex strength due to the larger induced AoA, which leads to the smaller magnitude of the negative lift in the upstroke.   This explanation echoes the observations of the weakened vortices generating the smaller negative vortex force on the lower surface (see Sections IV and V).

However, the quasi-steady lifting line model fails to predict the peak of $\Delta Cl(t)$ in the downstroke revealed by DNS.   Furthermore, the peak value and phase of $\Delta Cl(t)$ in the upstroke predicted by this model are different from those given by DNS.   It is not surprising since this simple linear model does not incorporate the non-linear vortex force associated with the leading-edge vortices on the upper and lower surfaces in both the downstroke and upstroke.   The classical lifting line model based on the Kutta-Joukowski theorem cannot correctly predict the amplitude and phase of the lift coefficient of a flapping wing in the highly unsteady and massively separated flows.

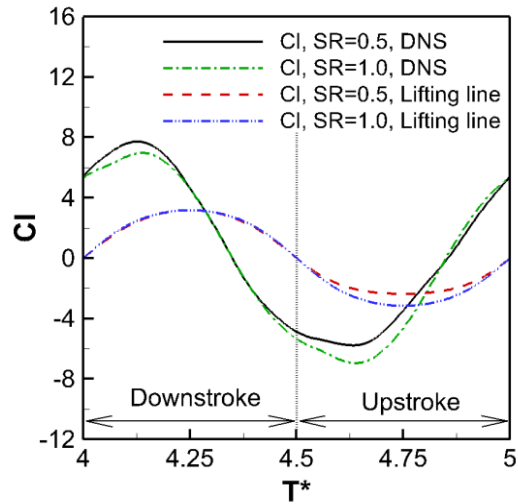

FIG. A1. The instantaneous lift coefficients $Cl(t)$ given by the quasi-steady lifting line model as a function of the non-dimensional time for the cases of SR = 0.5 and SR = 1.0, where the DNS results are plotted for comparison.

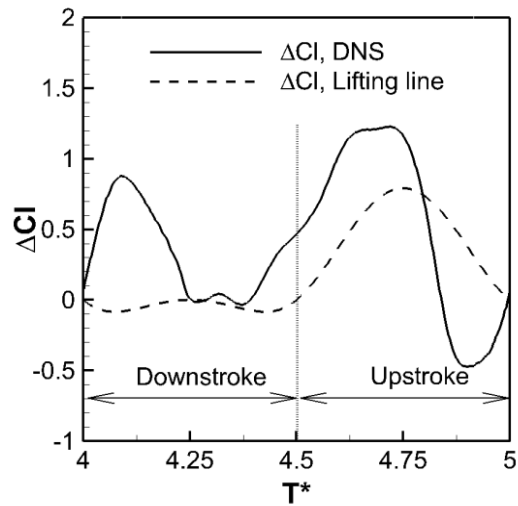

FIG. A2. The instantaneous increment $\Delta Cl(t)$ given by the quasi-steady lifting line model as a function of the non-dimensional time, where the DNS result is plotted for comparison.

# APPENDIX B. INDUCED DRAG CHANGE

The drag acting on the flapping wing in the parametric space $(SR, \phi, St, A^*, \alpha)$ is also calculated although the lift is the main theme in this work. Similarly, for comparisons between different cases, the drag coefficient based on $S(t)$ and drag coefficient based on $S_{max}$ are defined as $Cd = F_x(t)/0.5\rho U_\infty^2 S(t)$ and $Cd_{S\_max} = F_x(t)/0.5\rho U_\infty^2 S_{max}$, respectively, where $F_x$ is the drag, $S_{max}$ is the maximum wing area during a flapping cycle, and $S(t)$ is the instantaneous wing area. Accordingly, the increments of the drag coefficients are defined as $\Delta Cd = Cd - Cd(SR=1)$ and $\Delta Cd_{S\_max} = Cd_{S\_max} - Cd_{S\_max}(SR=1)$, respectively. Figure B1 shows the time histories of $Cd$, $Cd_{S\_max}$, $\Delta Cd$ and $\Delta Cd_{S\_max}$ in one period in the cases of $SR = 0.5$ and $SR = 1.0$ for $\phi = \pi/2$, $St = 0.3$, $A^* = 0.25$, and $\alpha = 0^o$. It is found that the drag coefficients $Cd$ and $Cd_{S\_max}$ for $SR = 0.5$ are larger than those for $SR = 1.0$. The increments $\Delta Cd$ and $\Delta Cd_{S\_max}$ are positive in the most portion of one period.

A question is whether the increment of the drag is related to that of the lift. To illustrate this point, Figure B2 shows the relationship between the time-averaged increments of the drag and lift coefficients $<\Delta Cd>_T$ and $<\Delta Cl>_T$ when $SR$ varies from 0.5 to 1.0 for $\phi = \pi/2$, $St = 0.3$, $A^* = 0.25$ and $\alpha = 0°$. There is reasonable correlation between $<\Delta Cd>_T$ and $<\Delta Cl>_T$, indicating that the drag increase is approximately proportional to the lift increase. In the light of the classical aerodynamics theory of a finite wing, it is implied is that the induced drag is increased as a byproduct of the lift enhancement through the altered vortical structures by stretching and retracting wingspan in flapping flight. This provides another evidence for the observation that the lift of the flapping wing is significantly enhanced by dynamically changing wingspan via the fluid-mechanic mechanisms besides the effect of changing the wing area.

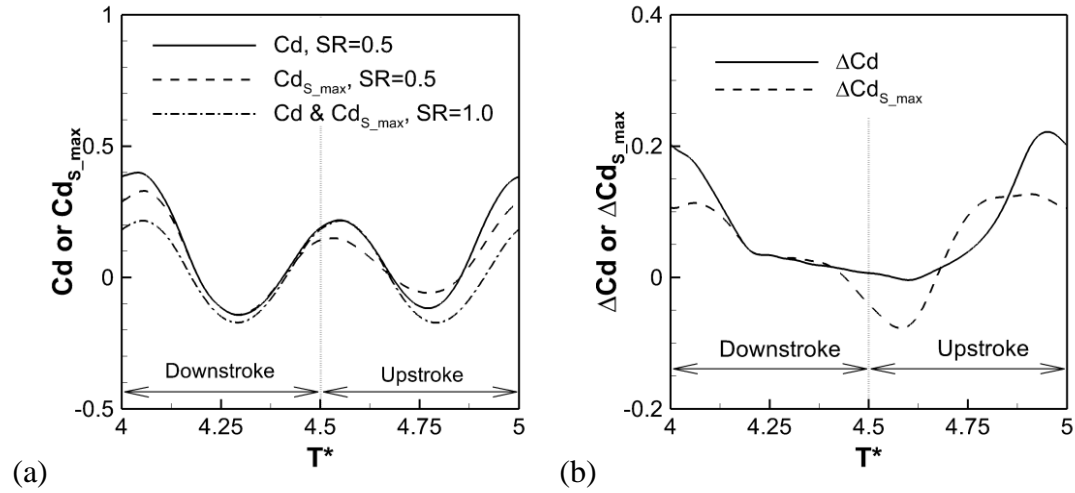

FIG. B1. The time histories of (a) $Cd$ and $Cd_{S\_max}$, and (b) $\Delta Cd$ and $\Delta Cd_{S\_max}$ in the cases of $SR = 0.5$ and $SR = 1.0$ for $\phi = \pi/2$, $St = 0.3$, $\alpha = 0^o$ and $A^* = 0.25$.

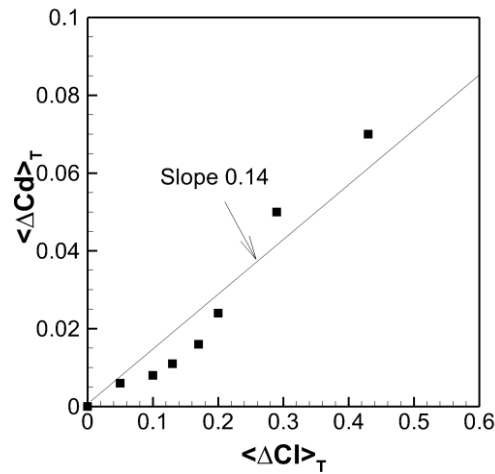

FIG. B2. The correlation between the time-averaged increments of the drag and lift coefficients $<\Delta Cd>_T$ and $<\Delta Cl>_T$ when $SR$ varies from 0.5 to 1.0 for $\phi = \pi/2$, $St = 0.3$, $\alpha = 0^o$ and $A^* = 0.25$.